\documentclass[]{aa}
\usepackage{natbib}
\usepackage{graphicx}
\usepackage{epsfig,rotate}

\bibpunct{(}{)}{;}{a}{}{,}

\newcommand{\be} {\begin{equation}}
\newcommand{\ee} {\end{equation}}
\newcommand{\anar}{{\it A\&AR, }}
\newcommand{\ana}{{\it A\&A, }}

\def\witchbox#1#2#3{\hbox{$\mathchar"#1#2#3$}}
\def\leqsim{\mathrel{\rlap{\lower3pt\witchbox218}\raise2pt\witchbox13C}}
\def\geqsim{\mathrel{\rlap{\lower3pt\witchbox218}\raise2pt\witchbox13E}}

\begin{document}

\title{Orbital circularisation of white dwarfs and
the formation of gravitational radiation sources in
star clusters containing an intermediate mass black hole}

\author{{\bf 
P. B. Ivanov$^{ 1,2{\large \star}}$,  J.C.B. Papaloizou$^{ 1}$}}


\offprints{ P. B. Ivanov}

\institute{  1-Department of Applied Mathematics and Theoretical Physics, CMS,
University of Cambridge, Wilberforce Road, Cambridge, CB3 0WA, UK \\
 2-Astro Space Center, P. N. Lebedev Physical Institute,
  Profsouyznaya St., 84/32, Moscow, Russia 
 \\ \email{pbi20@damtp.cam.ac.uk}}

\date{Accepted; Received; in original form;}

\label{firstpage}

\abstract
{}
{We consider how tight binaries consisting of a super-massive black hole
of mass $M=10^{3}-10^{4}M_{\odot}$ and a white dwarf in quasi-circular orbit
can be formed in a globular cluster. 
We point out that
a major fraction of white dwarfs tidally captured  by the black hole
may be destroyed by tidal  inflation during ongoing 
tidal circularisation, and therefore the formation  
of  tight binaries is inhibited. However some  fraction
may survive
tidal circularisation  through  being spun up 
to high 
rotation rates. Then  the  rates of energy loss through 
 gravitational wave  emission induced by   tidally excited pulsation
modes   and dissipation through non linear effects
may compete with the  rate of increase of  pulsation
energy due to dynamic tides. The semi-major axes of these white 
dwarfs are decreased by tidal interaction below a  'critical' value
where dynamic tides decrease in  effectiveness because pulsation modes retain 
phase coherence between successive pericentre passages. 
}
{We estimate the rate of  formation of
 such circularising 
white dwarfs within  a simple framework,  
 modelling them  as $n=1.5$ polytropes
and assuming that results obtained  from the tidal theory 
for slow rotators  can be extrapolated to 
fast rotators.}
{ We estimate the total capture rate 
as $\sim \dot N\sim 2.5\cdot 10^{-8}M_{4}^{1.3}r_{0.1}^{-2.1}yr^{-1}$, where
$M_{4}=M/10^4M_{\odot}$ and $r_{0.1}$ is the radius of influence
of the black hole expressed in units $0.1pc$. 
We find that the formation  rate of tight pairs 
is approximately 10 times
smaller than the total  capture rate, for typical parameters
of the problem. This result is used to estimate the probability
of detection of gravitational waves coming from   such tight binaries
by LISA.}
{  We conclude that LISA may  detect
such binaries provided that the fraction of globular clusters
containing  black holes in the mass range of interest
is substantial and that the dispersion velocity
of the cluster stars near the radius of influence of the black hole 
exceeds $\sim 20km/s$.}

\keywords{Black hole physics-Globular Clusters-White dwarfs-Tidal Circularisation}

\authorrunning{  Ivanov \&  Papaloizou}
\titlerunning{
 Gravitational radiation  from 
 clusters  with a 
 black hole  
 }

\maketitle

\section{Introduction }
\noindent

There are some observational indications  and theoretical suggestions  
  that favour  the  presence of black holes in the mass range 
$\sim 10^{2}-10^{4}M_{\odot}$ in the centres of globular clusters. The
observational arguments supporting this hypothesis   relate to
kinematical phenomena observed in the centres of some globular clusters
(e.g.  Gebhardt et al 2000, Gebhardt, Rich, $\&$ Ho 2002) and the presence
of X-ray sources not associated with the central nuclei
in certain galaxies (e.g. Fabbiano 1989, Matsumoto et al 2001, Ghosh
et al 2006 and references therein). There are also some theoretical 
models of the formation of such systems (e.g. Miller $\&$ Hamilton 2002).
A review of the observational and theoretical aspects of this problem
has been recently given by van der Marel 2004.

In this Paper we assume that there is a black hole of mass 
$M \sim 10^{3}-10^{4}M_{\odot}$ in a star 
cluster and estimate the rate of capture 
of white dwarfs of mass $m$ by the black hole. The capture rate is always
determined  by the interplay of  two processes. These are
the effect of distant two body 
gravitational encounters changing the orbital angular momenta of the
stars and  an interaction associated with 
the presence of the black hole
which removes the orbital energy of a star and is effective only 
when the orbital angular momentum  is sufficiently small. 
This type of interaction
 may result either through tidal interactions or by the emission 
of gravitational waves induced by the stellar orbital motion. 

The efficiency of tidal interactions is determined by the  ratio of orbital pericentre
distance to the tidal radius - the latter being the distance
from the black hole below which  significant disruption of the star
through mass loss induced by tides occurs.
On the other hand, the efficiency of 
 orbital energy loss due to gravitational wave
emission
is determined by the ratio of the
orbital pericentre distance to the gravitational radius of the black hole.

Since the tidal radii corresponding to white dwarfs
 for the range of black hole masses considered here
are larger than their gravitational radii, 
 orbital energy is 
changed mainly through the action of dynamical tides. 
This is  in contrast to the  case of  the  more 
massive black holes residing in galactic centres where emission of 
gravitational waves is
more effective for changing the orbital energy of
white dwarfs (e.g. Ivanov 2002, Freitag 
2003 and references therein).

Since the relative contribution  
of two body gravitational encounters to the orbital evolution
decreases very sharply  for small angular momenta, a star  with sufficiently 
small angular momentum loses orbital energy through  tidal interaction
while the orbital angular momentum remains approximately unchanged.
The latter occurs because the star cannot store a significant amount of angular
momentum in comparison to the orbit.
Accordingly, the orbital eccentricity  decreases
during this process which will be referred  to  as 
orbital circularisation. 

As a result of this process a tight
quasi-circular orbit around a black hole may be formed. A white
dwarf on such an orbit can emit gravitational radiation in the
frequency band of order of $10^{-2}Hz$ which is the
most favoured for the planned LISA space borne gravitational wave
antenna.  In principal  this is able to detect gravitational waves
with dimensionless amplitude as small as $10^{-24}$, for an 
observational time of one year. That means that the  presence of such 
a white dwarf 
can be detected from distances of order of $\sim 10^{3}Mpc$. 
Taking into account the fact  that globular clusters form a very abundant 
population of cosmic objects such systems may contribute significantly
to the budget of sources of gravitational radiation available for LISA.

There is one principal obstacle inhibiting formation of 
a binary pair consisting of a black hole of mass $M$ and  a white
dwarf of mass $m$ in a tight orbit around it.   Because it is produced
through tidal interaction, its semi-major
axis will be close to the tidal radius. There the ratio of orbital binding 
energy to gravitational   energy of the white dwarf is of order
of $(M/m)^{2/3} \gg 1$. Thus, for  
such an orbit to be produced,  an amount of energy  far exceeding the internal binding energy of
the white dwarf must be removed from it. When tides are
effective, orbital energy is transferred to pulsation
modes and thence  to the  internal energy of the star.  
Therefore, without an effective energy loss
mechanism,  the white dwarf could be easily unbound and thus
destroyed by the  tidal 
input of energy or tidal inflation. 

Here we propose and discuss such a
mechanism, which may, in principle, allow unbinding due to tidal energy input to be circumvented for
a range of  orbital parameters of the star. The operation of this mechanism
depends on the interplay of several  factors influencing the orbital
evolution of star which we introduce below.

The character of orbital evolution under the influence
of tides is mainly determined by three factors: 1) the time scale for
decay of the stellar pulsations excited by tidal interaction, 2) the orbital
parameters of the star, and 3) the rotation of the star.

Let us first consider the decay of pulsations in 
 a non-rotating star. In this paper we assume that decay of 
stellar pulsations is caused both  by the  emission of gravitational waves 
that occurs because of the time-dependent
density perturbations associated with them,  
and also by the dissipation of pulsation energy leading to its conversion 
into internal energy of the star.
The latter decay channel is assumed to result through non-linear effects.
Since its properties are very poorly understood at the present time,
we shall consider the corresponding dissipation time scale to be  a free
parameter. But we shall assume that it is larger
than the orbital period of the star. However, this is not an essential
assumption for orbital evolution at sufficiently large
semi-major axes, see below.

Initially the stellar orbit will be highly eccentric with a tidal interaction
that excites stellar pulsations
occurring impulsively at pericentre passage (e.g. Lai 1997, Ivanov $\&$ Papaloizou 2004 hereafter IP).
For long decay time scales,
pulsations will always be present in the star, and 
as a result of  every periastron passage, a new perturbation excited by tides 
is added.

 When orbital semi-major axis
is sufficiently large or the orbital period sufficiently long,
 it has been established that tidally induced changes to it cause the phase correlation between 
 preexisting pulsations and  freshly excited ones to be lost.
  Then, both the energy content
of excited pulsation modes and  the orbital energy of the
star evolve in a stochastic manner (eg. Kochanek 1992, Kosovichev $\&$
Novikov 1992, Mardling 1995). Under this evolution, the mode energy and the orbital
binding energy of the star grow on average, with a part of the mode
energy  being  transferred to the internal energy of the star. Therefore the
stochastic exchange of energy between the orbit and  stellar 
pulsations leads to a decrease of the orbital semi-major axis and
period as well as an increase of the internal energy of the star. 
    
However, once the orbital period is sufficiently short,
or  the semi-major axis is below a critical value,  
pulsation modes can maintain phase coherence
between successive pericentre passages, stochastic evolution ceases
and  dynamic tides are  expected to become less efficient. 
At this point, the tidal 
evolution rate is determined by the natural decay timescales
of the pulsation modes. A steady pulsation energy
typical of that induced through one pericentre passage may be maintained, rather than the
growth that occurs through the  cumulative effect
of mode energy inputs proceeding over many pericentre passages when the
 semi-major axis is large.

An important issue is whether the internal energy added to the
star during the phase of stochastic evolution is enough to cause 
its destruction.
For a non-rotating star the internal energy obtained by the star 
when the critical semi-major axis is reached is larger than the stellar
gravitational energy. Therefore, such a star could be disrupted by tidal 
heating before the critical semi-major axis is reached.
  
However, during orbital
circularisation orbital angular momentum is also transferred to the
star  until an equilibrium rotation rate is attained.
This requires a rotation rate corresponding to 
corotation at periastron or faster. 
Thus the star can be spun up to high rotation
rates. When measured in terms of
the amount of energy input per periastron passage,
the efficiency of tidal interaction is  minimised for such  a 
rotating star. 
See for example figure 8 of  Lai 1997 and also
IP. 
Accordingly the time scale of circularisation is increased.

\begin{figure}
\begin{center}
\includegraphics[width=9cm,angle=0]{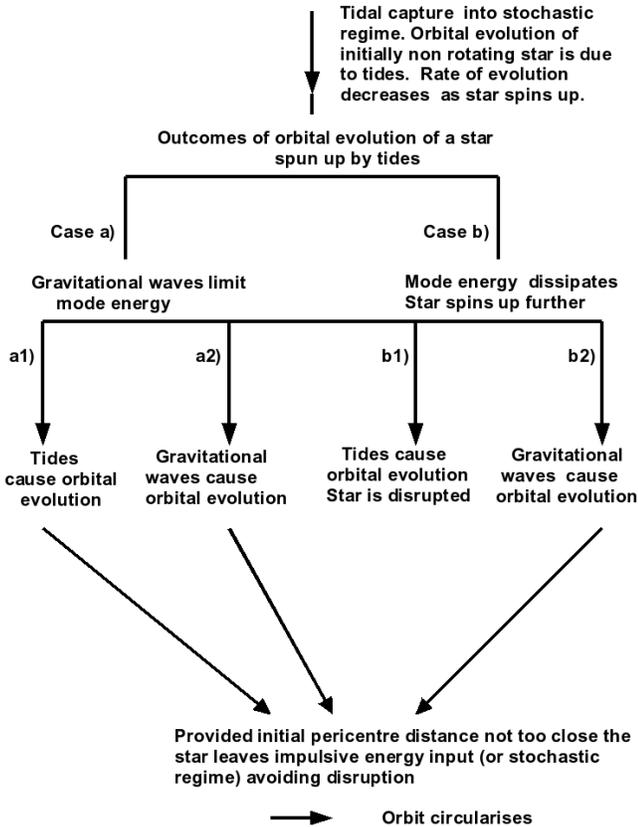}
\end{center}
\caption{A schematic illustration of the possible orbital evolutionary paths
of a white dwarf that is tidally captured by an intermediate mass black hole. 
The star is initially scattered into a highly elongated trajectory that evolves
due to impulsive energy and angular momentum exchanges between the
star and orbit that occur every pericentre passage, the pericentre
distance remaining very nearly fixed. As a result of these,
the internal energy and angular momentum increase. However, this process is slowed
as the star spins up and under favourable conditions gravitational radiation can become
important so that it controls the orbital evolution and/or damps the excited pulsation modes.
The latter case a) leading to cases a1) and a2) occurs where other non linear processes are
ineffective at dissipating the pulsations. Case b) leading to cases b1) and b2) occurs
when such processes are more effective. For both cases a) and b) the star may survive the orbital
evolution if the tidal capture occurs at sufficiently large pericentre distance.}
\label{fig01}
\end{figure}

In this situation, when dissipation of the mode energy 
as a result of non linear effects is not effective, 
the time scale for transmitting orbital
energy to the pulsation modes  may be 
larger than the time scale for removal
of the pulsation energy  through gravitational
waves emitted because of the  time-dependent perturbation of the star.
In this situation the white dwarf may reach the critical semi-major
axis without internal dissipation
causing tidal inflation
because of {\it cooling} by emission of gravitational waves.
In the opposite limit of effective mode dissipation and internal heating, the rapidly rotating
white dwarf may attain the critical semi-major axis
through the emission  of gravitational waves induced by orbital motion
rather than through tidal interaction.

Thus taking into account the reduction
in effectiveness of the tidal 
energy transfer brought about by the 
effect of stellar  rotation, we find that for sufficiently
large orbital angular momenta
the semi-major axis can be decreased
below the critical one without significant heating of the star.

After the critical semi-major axis is reached, again, because of the reduced
efficiency of the tidal interaction, orbital
evolution is governed by the  emission of gravitational 
waves determined by the orbital motion of the white 
dwarf. This process can further reduce the orbital
semi-major axis and lead to formation of a tight
quasi-circular orbit.

We formulate the criterion for white dwarf 'survival' treating
dynamic tides within the  framework of the simplest possible model
of tidal interactions. We assume that the internal structure
of a white dwarf is the same as that of a  $n=1.5$ polytrope.
We also assume that the results obtained   from the theory of dynamic
tides in slowly rotating stars can be extrapolated  to
high rotation rates for the purpose  of making
approximate estimates. 
Based
on these assumptions we estimate the formation rate of  tight
pairs which turns out to be an order of magnitude smaller than
the total rate of tidal capture, for typical parameters of the
problem. This allows us to obtain an estimate of 
probability of detection of such sources of gravitational waves
by LISA. We conclude that LISA could, in principal, detect such
a source provided that there is a significant fraction of
globular clusters containing black holes with masses 
$\sim 10^{3}-10^{4}M_{\odot}$, and with  stellar velocity dispersions
 in their innermost regions 
 exceeding $\sim 20 km/s$.

There are a  number of different possibilities and branchings associated 
with the orbital evolution
subsequent to tidal capture. Each of these requires consideration of several physical
processes and is described by a number of algebraic expressions. In order to  clarify 
the situation, we provide a more
transparent summary of the proposed paths to a circularised orbit with
disruption of the star avoided.
In Figure \ref{fig01}
we give a diagrammatic  illustration of the possible orbital evolutionary paths
that can be taken by 
a white dwarf subsequent to tidal capture by an intermediate mass black hole. 
Initially, impulsive energy and angular momentum exchanges between the  
orbit  and star that occur every pericentre passage spin it up and excite modes of oscillation. 
As a result the rate of tidal evolution of the orbit decreases
such that  gravitational radiation can become more
important. This may also be important for damping the oscillation modes.
If the tidal capture occurs at sufficiently large pericentre distance, 
the importance of gravitational radiation during the orbital and pulsation mode evolution
may allow the star to survive the regime of impulsive energy input
at pericentre passage and become circularised without disruption and
be a potential LISA source.

The plan of the paper, in which the above phenomena are considered in
more detail, is as follows.
In Section  \ref{sec1} we describe the model of the stellar cluster
we use and the effect of distant two body encounters.
In Section \ref{sec2}  we go on to discuss the
tidal interaction of a white dwarf with the central black hole and
pulsation mode excitation through dynamic tides.
Then in Section \ref{sec3} we consider the way in which
gravitational wave emission/non linear effects  
may lead to  pulsation mode amplitude limitation.
In Section \ref{sec4}  we   go on to explore the
conditions for safe circularisation or tidal disruption of the white dwarf.
In Section \ref{sec5}  we  discuss
the rate of tidal capture of white dwarfs  by the black hole 
and estimate the fraction that  can circularise safely, with technical
details being relegated to an Appendix.
In Section \ref{sec6}  we estimate
the probability of detection of   this source of gravitational
radiation. Finally in 
Section \ref{sec7}  we discuss our results.


\section{The stellar system}\label{sec1}
\noindent
 
\subsection{Density distribution}

In order to discuss the rate of  orbital circularisation for
white dwarfs it is first  
necessary  to consider 
 the properties of the star cluster  in which they
are situated. This is because  dynamical interaction with the stars
in  the system  is responsible for  the production
of orbits for which tidal interaction with the central black
hole becomes significant.

The radius of influence of the central
black hole, $r_{a}$, is defined as the radius
within which the total mass of the stars  is 
the same as the mass of the central black hole: $M_{st}(r_{a})=M$.
As has been noted by many authors, when $r < r_{a},$ a 'cusp'
in the stellar  distribution of stars is formed with a density distribution
such that
\be \rho \propto r^{-(3/2+p)}, \label{e1} \ee
where the parameter $p$ is typically  small. 
Two values that have been suggested  through theoretical
considerations  are  $p=0$ 
(Young, 1980) and
$p=1/4$ (Gurevich, 1964, Bahcall $\&$ Wolf 1976). 
Recently simulations by Baumgardt, Makino $\&$ Ebisuzaki
2004a, and  Baumgardt, Makino $\&$ Ebisuzaki 2004b 
have indicated that for the case of 
globular clusters  
having a  stellar population with a single mass,
$p \approx 1/4,$  while for  systems with
a realistic multi-mass stellar population, $p \approx 0.05,$
with $p \approx 1/4$ for the most  massive 
stars. 

In this paper we  make only  approximate estimates of the
tidal circularisation rate. Thus 
for simplicity, we  have adopted
$p=0$  for all components of the stellar system. 
We note that if a steeper cusp profile were to be adopted,
the number of tidally captured stars  at any particular time would increase  by a
numerical factor of order  unity.

\subsection{Phase space distribution } 
Making the usual assumption that the phase space distribution function for  
'typical' stars 
 depends on the binding energy per unit mass only
 and the distribution of white dwarfs  is the same as
that of  typical stars, 
 it can be easily shown from 
(\ref{e1}) with $p=0$
that the distribution  function for  the white dwarfs
   over 
 binding  energy per unit mass
 and specific angular momentum, $N^{0}_{wd}(E,J)$,
is
\be N^{0}_{wd}(E,J) =2^{-3/2}\delta \left({M\over m r_{a}^{3}}\right)
({GMr_{a}})^{-3/2}P_{orb}(E)J, \label{e2}\ee
where $M$ and $m$ are the masses of the black hole and of a
typical star, and we assume that $m \approx M_{\odot}$. 
$\delta \sim 0.1$ is the number fraction of white
dwarfs in the cusp, $E$ and $J$ are, respectively, 
the binding energy per unit mass and the
specific angular momentum of a star. We adopt the convention that the 
binding energy has positive values for the gravitationally bound stars
considered in this paper.  
\be P_{orb}(E)={\pi \over \sqrt 2}{GM\over E^{3/2}} \label{e3}\ee
is the orbital period of a star with binding energy per unit mass  $E$.

\subsection{Loss cone}
The assumption  of isotropy of the phase space distribution function
is applicable  only for estimates of quantities
determined by the bulk of the stars, such as e.g.  quantities
characterising distant gravitational interactions, see equation
(\ref{e5}) below.
It becomes invalid
for stars with sufficiently low orbital angular momentum which
can either be tidally disrupted or
directly captured by the black hole.
The presence of the
black hole thus leads to the  formation of a 'loss cone' 
such that equation (\ref{e2})
over-estimates the number of stars having low angular momenta.
Assuming that  stars with sufficiently
small specific angular momenta $J < J_{cone}$ are absent, 
the presence of the loss cone may be easily
accounted for by a correction factor (e.g. Lightman $\&$ Shapiro 1977)
such that the
distribution function of the white dwarfs is modified to become
\be N_{wd}={\log \Theta\over \Lambda_{1}}N^{0}_{wd}, \label{e4}\ee
where  $\Theta =J/J_{cone}$, $\Lambda_{1}=\ln (J_{circ}(E)/J_{cone})$,
and $J_{circ}$ is the specific angular momentum corresponding to a circular
orbit with a given value of the binding energy per unit mass $E.$ 
Typically, $\Lambda_{1}\sim 8$.

\subsection{Distant encounters}
A star orbiting around the black hole changes its energy and
angular momentum due to distant gravitational interactions with
other stars in the cusp. These result in a random walk
in phase space.  The stochastic change of the specific angular
momentum per  orbital period may be characterised by its
dispersion, $(\Delta j_{2})^{2}$, which can be written in the form (e.g., 
Ivanov, 2002):
\be (\Delta j_{2})^{2}={29\pi \over 20} 
\Lambda_{2} q(GMr_{a}){\cal E}^{-5/2},
\label{e5}\ee
where $q=m/M$ is the mass ratio, $\Lambda_{2}\approx \ln( 0.5 q^{-1})$ 
and we use the dimensionless energy variable $\cal E$ defined  through
${\cal E}=Er_{a}/(GM)$.\footnote{In what follows it is convenient to
represent different energy scales in dimensionless form. There are two
natural  units  of energy per unit mass in our problem: the first  associated the stellar
cusp - $GM/r_{a}$ and the second  associated with the internal structure of star itself -
$Gm/r_{wd}$. For the dimensionless quantities expressed in terms of the
first  unit we will use the calligraphic style while the quantities
expressed in terms of the second unit will be denoted by tilde.}


\section{Tidal interaction of a white dwarf with the central black hole}\label{sec2}
\noindent

\subsection{White dwarf model}

In order to make order of magnitude estimates 
we use the simplest possible model  for the
white dwarf. We assume zero temperature so that
the pressure is due to completely degenerate  electrons.   
In this case the equation of state is
baratropic and 
gravity or $g$  modes are absent.
Furthermore  we approximate the structure of the white dwarf as  
that  of a $n=1.5$ polytrope.   In addition to  the mass $m$ and radius $R_{wd}$
we introduce the two parameters $\mu= m/M_{\odot}$
and $\lambda = R_{wd}/(0.01R_{\odot})\approx
R_{wd}/(7\cdot 10^{8}cm)$. The effectiveness of tidal  interactions
depends significantly on  the mean density of the white dwarf.
To characterise  low and high
density  cases we consider two 
sets of values for  $\mu $ and $\lambda.$ 
The 'low density' case will be characterised by parameters appropriate for
a 'typical' white dwarf   for which $\mu =0.6 $ and $\lambda =1.4$, and
for the  'high density' case we  use  parameters corresponding
to Sirius B   for which $\mu \approx 1$ and $\lambda \approx 0.84$.     


\subsection{Tidal perturbations and orbital circularisation}

For a baratropic star only fundamental
$(f),$ pressure $(p)$ and inertial modes can be excited  by
tidal interaction. Since $p$-modes have large eigen-frequencies and  inertial
modes are significant only for rather large orbital angular momenta
(Papaloizou $\&$ Ivanov 2005),
their respective contributions to the tidal exchange of energy and
angular momentum are small compared to the contribution of the
$f$-modes for the range of the orbital parameters  of
interest, so they are not considered further.

To make our estimates, we use  the linear theory of  tidal
perturbations induced in a slowly rotating star and consider
only  the leading order quadrupole response. The assumption 
of slow rotation is  certainly valid in the  early
stages of orbital evolution due to tides.  In this regime  orbital changes
induced by distant gravitational interactions and  tidal  effects compete
with each other.  Then we may assume that the star is non-rotating
when calculating the rate of tidal circularisation.

However,  the
star can be spun up  because of the effect of tides during 
orbital circularisation, 
achieving a considerable rotation rate  during the later stages.  
As we will see below this  may have important
implications for the outcome of the orbital evolution of the star. 
In particular, it may play a role in  determining
whether is it possible or not  for the star to  survive in
a tight quasi-circular orbit. Therefore,
it is directly  related to the problem  of formation of  sources
of gravitational radiation.  Unfortunately, a theory of tidal
perturbations of a fast rotating star is practically  undeveloped
in the present time.   Accordingly, we   shall assume that the 
results obtained  from the theory of slowly rotating stars can
be extrapolated to the case of fast rotation  for the purposes of making
order of magnitude estimates.

During the early stages of tidal circularisation, the  orbit of the star 
is highly eccentric. In this case  tidal 
interactions are only important near pericentre. Because of this, in order to  
calculate the energy and angular
momentum exchange between the orbit and the modes of pulsation of
the star
it is possible to consider it as undergoing a sequence of successive
flybys of the black hole each of which produces an  impulsive change. 
This formulation of the problem allows us to use 
the theory of tidal excitation and dissipation of
the fundamental modes in a baratropic star moving on a highly
eccentric or parabolic orbit developed elsewhere
(see e.g. Press \& Teukolsky 1977 hereafter PT, 
Lai 1997, IP). 
This theory is based on consideration
of a single pericentre passage where an amount of energy
$\Delta E$ and an amount of angular 
momentum $\Delta L$ is transferred from the orbit to the star. 
These quantities depend on the  amplitudes and phases of the pulsation modes
before the  passage and
on the radius of  pericentre, $r_{p}$.

Let us temporarily assume that the amplitude of the pulsation mode
before  pericentre  passage is negligible and the stellar model
and the angular velocity of the star are fixed.
In this case, it can be shown that the quantities $\Delta E$ and
$\Delta L$ are 
determined by the value of dimensionless parameter (PT) 
\be \eta=\sqrt {{m\over M}\left({r_{p}\over R_{wd}}\right)^{3}}, 
\label{e6n}\ee
where $r_{p}$ is the pericentre distance. These quantities can 
be naturally  considered as being made up of two parts: 1) a quasi-static part 
determined by irreversible loss of  total energy of the system
during the pericentre passage and  2) a dynamic part (due to the so-called 
dynamic tide) determined by excitation
of the normal modes of the star.

\subsection{Quasi-static tides}
The quasi-static part of the energy transfer $\Delta E_{q-s}$,
can be written in the form: $\Delta E_{q-s} \approx \gamma
E_{*}/\eta^{5}$, where $E_{*}=Gm^{2}/R_{wd}$ is some 'reference'
energy associated with the star and the dimensionless coefficient
$\gamma $ is determined by the dissipative processes  
 operating in the star.  An explicit expression for $\gamma $ 
can be found in IP. During the flyby
 energy loss occurs  as a result of 
viscous dissipation in the bulk of
the white dwarf and in its convective envelope and also
due to emission of gravitational waves. Since the coefficient
of dynamic viscosity in the bulk of the white dwarf appears
to be very small ($\sim 10^{4}-10^{6}$ in cgs units, e.g. Chugunov 
$\&$ Yakovlev 2005 and references therein) and convective 
envelopes  of white dwarfs also have a rather small estimated
  turbulent viscosity and a very small relative mass, their
respective contributions to the coefficient $\gamma $ are
quite small. The main contribution appears to  come from
 the  emission of gravitational waves generated by time dependent
perturbation of the white dwarf (see Osaki
$\&$ Hansen 1973, hereafter OH). However, a simple estimate
shows that the value of $\gamma $ determined by this process
 leads to a value of $\Delta E_{q-s}$ much smaller than
that  associated with  dynamic tides, for relevant values 
of $\eta $.  Therefore, the overall
contribution of  quasi-static tides appears to be negligible
and will not be considered further.  

\subsection{Dynamic tides}
Now let us consider dynamic tides associated with
the fundamental quadrupole mode of pulsation. As was shown by IP
for the case of a slowly rotating star, 
the quadrupole mode propagating in the direction of orbital motion
(the so-called prograde mode) determines the exchange
of energy between the orbit and the star provided that
the angular frequency of the star, $\Omega_{r}$, is
smaller that its 'equilibrium' value 
$\Omega_{eq}\approx (2+\ln \eta ) \Omega_{*}/\eta$,
where $\Omega_{*}=\sqrt{Gm/R_{wd}^{3}}$ is a 'reference' 
frequency associated with the star. As it
will be shown later, in our problem a typical value
of $\eta \sim 4$, and the equilibrium angular frequency
$\Omega_{eq}\approx 0.8\Omega_{*}$ is formally larger than 
the angular frequency at rotational break-up of the star,
$\Omega_{br}\approx 0.5 \Omega_{*}$. Although the theory
leading to this conclusion is not valid at such high rotation
rates, it seems reasonable to assume that the energy exchange
is mainly determined by the prograde mode unless a rate of
rotation very close to $\Omega_{br}$ is reached. 
Considering excitation of only this
mode and assuming that the star is not
perturbed before pericentre passage, it easy to see 
from results given in IP that
the energy gained by the
star per unit of mass, $\Delta E_{T}$ can be expressed as
\be \Delta E_{T} \approx 0.7\phi^{-1}\Psi^{-1}{Gm\over R_{wd}},
\label{e6}\ee where we  use the fact that the eigenfrequency 
of the quadrupole $f$ mode  for
 a non-rotating n=1.5 polytrope $\omega_{f}\approx 1.445\Omega_{*}$
and  the value of dimensionless 'overlap' integral $Q_{f}\approx
0.5$ calculated in, e.g., Lee \& Ostriker 1986.   
The quantity $\phi(\eta )$ characterises the
strength of the tidal interaction for a non-rotating star
and $\Psi$ determines correction to the energy exchange due to
rotation, being unity for a non rotating star (see below).
Explicitly, we have
\be \phi(\eta)=\eta^{-1} \exp (2.74(\eta-1)), \label{e7}\ee
where
$\eta$ is given by equation (\ref{e6n}). When $\eta \sim 1-2,$ the
white dwarf is either tidally disrupted or stripped, and
the linear theory of dynamic tides cannot be used (see e.g. Frolov et
al. 1994; Ivanov
$\&$ Novikov 2001 and references therein). 
As it will be shown later, typically, $\eta
\sim 4-5 $ for  white dwarfs undergoing tidal circularisation.
Since the energy gain decreases exponentially with $\eta$ the use
of the linear theory seems to be justified in our case. 

\subsection{Direct capture or tidal disruption}

The characteristic   pericentre distance   below which tidal disruption is expected
 is determined as the distance where the tidal force acting on an
 element of the star near its surface is comparable to the
 gravitational  force  due to  the star itself. Thus it is given by 
\be r_{T}=\left({{M\over m}}\right)^{1/3} R_{wd}\approx 1.5\cdot 10^{10}\lambda
\mu^{-1/3}M_{4}^{1/3}
cm,
\label{e8} \ee
where $M_{4}=M/(10^{4}M_{\odot})$. Using this 
we can rewrite equation (\ref{e6n}) in another form
\be \eta =\left({J\over J_{T}}\right)^{3}, \label{e9}\ee
where $J_{T}=\sqrt {2GMr_{T}}$ is the specific orbital angular momentum
of an orbit with pericentre distance  equal to $r_{T}$. From equation 
(\ref{e9}) it follows that when $J\approx const$ $\eta \propto J^{3}$ is
also approximately conserved. 

The black hole mainly tidally  disrupts the white dwarfs
provided that its mass is sufficiently small.
 To see this we note that the specific orbital angular
momentum of  stars directly captured by the black hole   must be smaller
than $J_{cap}= 4GM/c .$
 Define the 'critical' mass $M_{crit}$  by the  condition
that $J_{cap}=J_{T}$ when $M=M_{crit}.$ One obtains
\be  M_{crit}=1.4\cdot 10^{4}\lambda^{3/2}\mu^{-1/2}M_{\odot}. \label{e10}\ee
 When $M < M_{crit}$, the
specific angular momentum $J_{T}$ exceeds  $J_{cap}$ so  it  
  determines the boundary of the loss cone. Thus
$J_{cone}\approx J_{T}$ and, accordingly, $\eta \approx \Theta^{3}$
( recall that  $\Theta $ is defined as $\Theta =J/J_{cone}$).

\subsection{ Effect of stellar rotation}

As indicated above, the rotation rate of the star has an
important effect in determining the strength of the tidal interaction.
From the results given in IP it follows that the corresponding
correction factor $\Psi$ entering in equation (\ref{e6}) has the form
\be \Psi \approx \exp {{4\sqrt 2\over 3} \tilde \Omega_{r}\eta},
\label{e11}\ee where $\tilde \Omega_{r}=\Omega_{r}/\Omega_{*}$,
$\Omega_{r}$ is the angular velocity of the star and we recall that
$\Omega_{*}=\sqrt {{Gm\over R_{wd}^{3}}}$ characterises a typical
frequency of stellar pulsation associated with the fundamental
mode. It is important to note that equation (\ref{e11}) has been formally
derived assuming that $\tilde \Omega_{r} \ll 1$.
However, we are going to use it later on to make our  
estimates implying that it is approximately valid when  
$\Omega_{r} < \Omega_{br}\approx 0.5 \Omega_{*}$.
Note that this assumption may be justified in part by results of
calculations of Managan (1986) who considered normal modes of
a fast rotating $n=1.5$ polytrope and 
of Clement 1989 who considered those of a fast rotating
$15M_{\odot}$ star. In both cases it was  
found that the eigen frequency of the 
prograde $f$ mode has an approximate linear dependence  
on the frequency of rotation until an angular frequency very
close to $\Omega_{br}$ is reached. The linear dependence
of the eigen frequency on the frequency of rotation appears to
be one of the most important factors leading to equation 
(\ref{e11}).

\subsection{ The effect of many pericentre passages and the criterion for
stochastic instability}\label{stocha}

Let us consider  a sequence of many pericentre flybys assuming that
the energy per unit mass contained in modes of oscillation, $E_{m}$, does not
decay significantly between successive pericentre passages. In this
case there is interference between the preexisting wave  perturbation 
in the star and the wave excited in the vicinity of pericentre as a
result of tidal perturbation. In this case, simple addition of
 (\ref{e6}) to $E_{m}$ after the pericentre passage to obtain
the new  $E_{m}$ needs
further justification. In this context it is important that the
orbital period $P_{orb}$ is changing with time as a result of tidal
interaction and loss of orbital energy due to emission of
gravitational waves
\footnote{Note that it is important to distinguish between the effect 
of emission of gravitational waves by perturbations of the white dwarf
and the effect of emission of gravitational waves due to the 
orbital motion of the star. The former effect is important as a 
major process of mode energy loss while the latter significantly influences the
orbital evolution and may provide a source of gravitational waves 
during the late stages of orbital evolution, see below.}, and therefore
the  the phase change  of the  oscillating mode generated
between two successive pericentre passages,  $ \Phi
=\omega_{f}P_{orb}$, is also changing with time. 
When the change of  $\Phi $ during one orbital period,
$\delta \Phi = \omega_{f}\delta P_{orb}$ is larger than 
$2\pi $, the mode phase at the time of the next pericentre  passage 
is essentially a random quantity, and accordingly, the mode 
energy  per unit mass $E_{m}$ 
evolves as a stochastic variable. In this case
we can add expression (\ref{e6}) to $E_{m}$ at 
each pericentre passage on average,
 assuming 
that all resulting expressions are valid in some statistical 
sense being averaged over many realisations of the stochastic 
process. The process of  stochastic evolution of the mode
energy will be refereed to as 'stochastic instability'.

A condition for the stochastic instability to operate can be
readily found from the condition $\omega_{f}|{dP_{orb}/ dE}|\Delta
E  > 2\pi$ and the expression (\ref{e3}) for the
orbital period in the form
\footnote{A more accurate expression has been derived in 
 IP. See also this paper for the references on numerical and analytical 
studies of the stochastic instability.}
\be a > a_{st}\approx {(GM)^{3/5}\over (3\omega_{f}\Delta E)^{2/5}}=
\lambda \mu^{-3/5}M_{4}^{3/5}(\Delta \tilde E)^{-2/5}10^{11}cm,
\label{e12}  \ee
where $a$ is the orbit semi-major axis,
\be \Delta \tilde E \equiv \Delta \tilde E_{T} + \Delta \tilde E_{GW}\equiv
\left({R_{wd}\over Gm}\right)(\Delta E_{T} + \Delta E_{GW}) \label{e13}\ee
is the dimensionless change of the orbital energy per unit mass, $\Delta E,$ in one
orbital period, and
$\Delta E_{GW} $ is the change of the orbital 
energy per unit of mass during one orbital period 
due to emission of gravitational waves, see equation (\ref{e21}) below
for an explicit expression. As follows from equation 
(\ref{e13}) the condition for the stochastic instability to set in
involves two contributions to the change of the orbital
energy: $\Delta E_{T} $ associated with tides and $\Delta E_{GW} $
determined by gravitational waves emitted as a result of orbital
motion. This differs from the case of tidally interacting 
stellar binaries where only the tidal contribution is normally
considered. In the case of stellar binaries, when the stochastic 
instability operates, both the orbital binding energy and the energy
of the mode of pulsation evolve in a stochastic manner (e.g. Mardling
1995).  The fact that $\Delta E_{GW} $ enters into equation 
(\ref{e13}) leads to the possibility of the stochastic instability
 operating in a situation where orbital evolution is mainly 
determined by emission of gravitational waves, i.e. $\Delta E_{GW} \gg
\Delta E_{T} $, see Section 5 and the Appendix. 
In this limit, our numerical investigations indicate that  
the orbital binding energy  can increase in a  way  that appears  to be fairly regular while the energy 
of the  pulsation mode still tends to increase   
in an erratic  manner. On the other hand, the mode energy ceases to increase when 
$a < \sim a_{st}$  and $\Delta E_{T}/\Delta E_{GW} > \sim 0.05$. We have checked 
that the condition  $\Delta E_{T}/\Delta E_{GW} > \sim 0.05$ is fulfilled for parameters typical for our 
problem.

It is interesting to note
that the condition (\ref {e12}) can be also derived from quite 
a  different approach based on the treatment of the tidal interaction
as occurring through  resonances between the mode frequency 
$\omega_{f}$ and the changing  orbital frequency $\Omega_{orb}$ of different 
order $n$, such that $n\Omega_{orb}=\omega_{f}$.  The
condition of overlap of these resonances given that $\Omega_{orb}$
is changing is  
$\delta n =\omega_{f}|{d \Omega_{orb}^{-1}/ dE}|\Delta E > 1.$ 
This is equivalent to the condition (\ref{e12}). 

\section{Gravitational wave emission/non-linear effects, characteristic time scales
 and  mode amplitude limitation}\label{sec3}

As we have mentioned above the most important linear decay channel of 
the fundamental mode appears to be through emission of 
gravitational waves. The corresponding decay time of the fundamental
mode has been calculated  by OH for two models of  iron 
white dwarfs. 
The results can be extended to the range of
masses and radii appropriate for  'standard' CO white 
dwarfs with help of the  expression for the gravitational
wave luminosity, $L_{GW}$, produced by the oscillating  mode  and the resulting  decay time
can be expressed as
\be t_{dec}\approx 1.5\cdot 10^{2}\lambda^{4}\mu^{-3}yr. \label{e14}\ee 
Note we define the decay time as $t_{dec}=E_{m}/L_{GW}$
which gives a value of $t_{dec}$ which is a factor of two smaller  compared 
to that of OH.

In this situation the mode energy can attain a large value
as a result of the cumulative effect of 
multiple tidal interactions occurring at pericentre passage.
At high amplitudes non linear effects may lead to additional
dissipation and  mode decay.
We characterise such effects by adopting a mode decay time scale
$t_{nl}$ which is a function of the mode
 energy. The non linear theory for these pulsation modes
is not yet adequately developed to provide a form for $t_{nl}$. Therefore,
we regard this as a free parameter and consider
two limiting cases: a) The time $t_{nl}$  is assumed to be large 
compared to  $t_{dec},$ being taken to be
comparable to $t_{ev},$ the {\it actual} time scale of orbital evolution
which is the smaller of the times required to change the
orbit significantly through tidal interaction  or to make it evolve as 
a result of the emission of gravitational waves induced
by that orbital motion, see below equation (\ref{eq18a}).
Assuming that all quantities characterising the mode of pulsation and the star
evolve on the time scale of order of $t_{ev}$, we estimate below that the condition
$t_{nl}\sim t_{ev}$ is equivalent to the condition that the rotational angular momentum
of the star is the same order of magnitude as the angular momentum corresponding to the
mode, see equation  (\ref{en1}) below. Thus, in this situation the degree of non-linearity 
adjusts in a way that does not allow the mode angular momentum to significantly exceed that in
the stellar rotation. 
b) The time $t_{nl}$
is assumed to be short compared to both  $t_{dec}$ and 
the orbital evolution time scale $t_{ev}.$ 
 However, in both cases   we shall assume that  $t_{nl}$ 
is much larger than the orbital period of the tidally interacting star.
But note that this is not strictly necessary during the first phase of evolution
when the semi-major axis is large ( see section \ref{2.1}  below)

\subsection{Balance of mode energy and angular momentum
 input due to tides and losses due to
gravitational radiation/non linear effects}\label{2.1}

Here we estimate the amount of  mode energy 
that is stored in the star when there is a balance between  
the build-up of mode energy due to
the stochastic instability and decay due to the emission of gravitational
waves or non linear effects. 
Provided this balance can occur with a stellar
mode angular momentum content that is small enough to avoid break up
of the star,
it  is possible for  it to 
avoid disruption by having to absorb  a large amount
of released orbital binding energy.
In a stationary state   
the mode energy per unit of mass can be estimated from the balance equation
\be E_{m}\left({1\over t_{dec}}+{1\over t_{nl}}\right) =
 {\Delta E_{T} \over P_{orb}} \label{e15} \ee
The  mode specific angular momentum, $L_{m}$, is related to $E_{m}$ by the
standard equation: 
\be L_{m}=2{E_{m}\over \omega_{f}}, \label{e16}\ee
where we take into account that tides mainly excite the quadrupole  mode
with the azimuthal mode  number $m=2$.

The rate of transfer of angular momentum to the star is
related to the rate of dissipation  of mode energy
(eg. Goldreich \& Nicholson 1989) and thus determined by
$t_{nl}.$ The rate of transfer of specific angular momentum  is
 $\dot L_{*}=L_{m}/t_{nl}$ and accordingly the
total specific angular momentum transferred to the star in a time 
interval $t_{ev}$  is
estimated to be
\be L_{*} \sim \left({t_{ev}\over t_{nl}}\right )L_{m}.
\label{en1}\ee 
Obviously, $L_{*}$ should not exceed the specific angular momentum of
the star in a state of uniform rotation with  angular frequency 
corresponding to  rotational break-up.

Estimating the moment of 
inertia of the star per unit of mass as $I \approx 0.2R_{wd}^{2}$ and 
using the standard relation $L_{*}=I\Omega_{*}\tilde \Omega_{r}$, 
from equations
(\ref{e15}), (\ref{e16}) and (\ref{en1}) we obtain
the associated dimensionless angular velocity of the star   
to be given by 
\be \tilde \Omega_{r}\approx 7{t_{dec}\over (t_{nl}+t_{dec})}\left ({t_{ev}\over
t_{T}}\right )\left({R_{*}E\over Gm}\right), \label{e17a}\ee
where we introduce the characteristic time
for tidal build-up of the mode energy
\be t_{T}\equiv {E\over \Delta E_{T}}P_{orb}. \label{e18}\ee
Note that $t_{T}$ so defined can formally be smaller than $P_{orb}$.
It is obvious that in that case it is the orbital period 
which plays the role of the characteristic tidal time scale.
Substituting equation (\ref{e6}) in 
(\ref{e18}) we can express the  tidal time scale in the form
\be t_{T}\approx \lambda \mu^{-1}
(\phi \Psi){\cal E}^{-1/2}r^{1/2}_{1}yr,
\label{e19}\ee 
where  $r_{1}=r_a/(1pc)$.
Note that since $(\phi \Psi) \gg 1$ typically we have $t_{T} \gg 1yr$. 

In general the time scale of orbital evolution
is given by
\be t_{ev}\approx \min{(t_{T}, \quad t_{GW})}, \label{eq18a}\ee
where $t_{GW}$ is the characteristic time of orbital evolution
due to  gravitational wave  emission explicitly given below,
see equation (\ref{e22}).
As  indicated below, for a white dwarf
spun up through tidal interaction,  gravitational
waves emitted as a consequence of orbital motion
are more effective than tides  at
inducing orbital evolution. In this case
$t_{ev}\sim t_{GW} \ll t_{T}.$

Recalling that the break up angular velocity
$\Omega_{br}\approx 0.5\Omega_{*}$, and assuming that
the criterion $\Omega_{r} < \Omega_{br}$ provides the condition for the
white dwarf to survive the circularisation process, 
we obtain a limitation on the orbital binding energy per unit mass 
of the white dwarf, during the phase of stochastic instability,
to be given by 
\be  E <  E_{dis}=7\cdot 10^{-2}{(t_{nl}+t_{dec})\over t_{dec}} 
\left({t_{T}\over t_{ev}}\right){Gm\over R_{wd}}. \label{e17}\ee
This equation leads to
a constraint that  the orbital binding energy of the white dwarf should
not be too high while mode energy input as implied by the existence of
stochastic instability is occurring.
The condition (\ref{e17}) can be simplified
when $t_{nl}$ is either large or small. 
For case a) 
when $t_{nl}\approx t_{ev}$ the condition
(\ref{e17}) can be written as
\be  {\cal E} < {\cal E}_{dis}\equiv \left({r_{a}\over GM}\right)E_{dis}=
7\cdot 10^{-2} {m\over M} {r_{a}\over R_{wd}}
{t_{T}\over t_{dec}}, \label{en2}\ee
where we recall that the dimensionless energy $\cal E$ is related 
to the orbital binding energy per unit mass  $E$ as 
${\cal E} =({r_{a}/GM})E$.
This condition  can be reformulated as setting
an upper bound on the allowed  mode energy per unit mass as 
\be E_{m} < 7\cdot 10^{-2} {Gm\over R_{wd}}. \label{e18a}\ee
Note  that in case a)  the angular momentum transferred to stellar rotation
is approximately equal to that associated with the mode of pulsation.
Therefore the condition  $t_{nl}\approx t_{ev}$ can be regarded
as a condition for equipartition of the angular momentum associated with the pulsation
mode and that associated with the stellar rotation.

In the opposite limit corresponding to case b) we have $t_{nl} \ll t_{dec}, t_{ev}.$
Then equation (\ref{e17}) gives
\be {\cal E} < {\cal E}_{dis}=7\cdot 10^{-2} {m\over M} {r_{a}\over R_{wd}}
{t_{T}\over t_{ev}}. \label{en3}\ee
In this case the energy put into internal energy of the star, $E_{int}$, through dissipation 
is much larger than the mode energy, $E_{m}$, and
we can express (\ref{en3}) as leading to  an upper bound for
$E_{int}$ in a form analogous to that given by
(\ref{e18a})
\be E_{int} < 7\cdot 10^{-2} {Gm\over R_{wd}}. \label{en4}\ee

We remark that this condition, obtained from considerations
of the stellar rotation, is equivalent to specifying a limit
on the allowed increase of the internal energy. 
 As almost all of the orbital binding energy is 
converted to internal energy of the star in case b), 
the same constraint would occur if the initial evolutionary phase involving stochastic instability
 were replaced by one in which the
 energy input into oscillations  during each pericentre passage
was dissipated before the next one. Thus, as indicated in section
 \ref{sec3} above, the condition that
$t_{nl}$ should be less than the orbital period is not  strictly necessary
during the initial evolutionary phase following tidal capture 
 (see also IP). 

Note too that we have  neglected the change of 
the white dwarf radius due to tidal
heating. Assuming this change is modest, we can take it into account
by considering the white dwarf to have a slightly smaller mean density
when discussing conditions for safe circularisation, see 
the next Section. 

\subsection{Effect of gravitational radiation on the orbit}\label{3.1}
The change of orbital energy  per unit mass
due to emission of gravitational waves generated by orbital motion
of the star 
can be obtained from results given by
Peters 1964. In the case of a  highly eccentric orbit,
the change per orbital period, which is mainly induced at pericentre
 may   be written
in the form
$$\hspace{-2.5cm} \Delta \tilde E_{GW}=3.8\cdot 10^{-3}\left({J\over
 J_{cap}}\right)^{-7}\left({c^{2} R_{wd}\over GM}\right)$$
\be \approx 8\cdot 10^{-4}\eta^{-7/3}
\lambda^{-5/2}\mu^{7/6}M_{4}^{4/3}, \label{e21} \ee
and the corresponding evolution time can be expressed in terms
of $t_{T}$ as
\be t_{GW}=\left({\Delta \tilde E_{T}\over \Delta \tilde E_{GW}}\right)t_{T}. 
\label{e22}\ee
Since the corresponding quantity induced by
tides, $\Delta E_{T},$ decreases with $\eta$ exponentially and
 $\Delta E_{GW}$ as an inverse power of  $\eta$, there is
a certain value of  $\eta$ where $t_{T}=t_{GW}$. At larger
values of $\eta $ the orbital evolution is mainly determined 
by emission of gravitational waves.   For
 non rotating stars (i.e. setting $\Psi=1$) this
value of $\eta $ is approximately equal to $\sim 5.3$
 for a 'high density' white 
dwarf and $\sim 5.6$ for a 'low density' white dwarf.
As we have mentioned
 above and   will discuss below, a typical value of $\eta $
for the stars undergoing tidal circularisation is of the order of
$\sim 4$ and always smaller than 
$\sim 5$. Taking into account that $t_{T}$    increases very rapidly
  with this parameter, we can conclude that tidal
interactions play the most significant role in the formation of a
flow of stars undergoing orbital circularisation.

On the other hand, the star is spun up by tides during 
ongoing orbital circularisation and
$t_{T}$ is increased as a result. Therefore, at a later stage,
the orbital evolution may be governed by emission of 
gravitational waves when the orbital binding energy is sufficiently 
large, depending on parameters of the white dwarf 
and of the star cluster.
Moreover, when the binding energy exceeds the value corresponding
to the onset of  stochastic instability, the tidal response
becomes quasi periodic and
ineffective   so that the orbital evolution is solely determined by the 
emission of gravitational waves.

\section{Conditions for safe circularisation or tidal disruption}\label{sec4}  
\begin{figure}
\begin{center}
\includegraphics[width=9cm,angle=0]{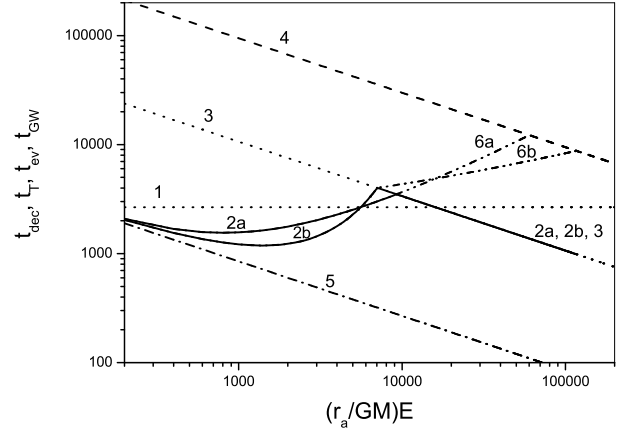}
\end{center}
\vspace{-1cm}
\caption{Characteristic time scales
years as functions of the dimensionless orbital energy $\cal E.$
Plotted here are  $t_{dec},$  the decay time of the
pulsation mode  due to emission of gravitational waves associated with
the stellar pulsation, $t_{ev},$ 
the time scale for  evolution of the orbit, $t_{T},$  the
time scale of  build-up of the mode energy  and 
$t_{GW},$ the time scale of orbital evolution due to emission of
gravitational waves  induced by orbital motion.   
The horizontal dotted line 1
represents $t_{dec}.$  The two solid curves 2a and 2b represent $t_{ev}$ 
calculated  assuming that
 the   stellar rotation 
frequency  is given by equation (\ref{e17a}).
 The curve 2a  taking on larger
values at small values of $\cal E$ represents the case $t_{nl}\approx t_{ev}$,
and the other  curve  represents the case $t_{nl}\ll t_{ev}$.
The  inclined
dotted  line 3 represents $t_{GW}$ which becomes equal to
$t_{ev}$ at the larger binding energies where
gravitational radiation controls the orbital evolution. 
The dashed and dot dashed
lines 4 and 5 represent $t_{T}$  calculated for $\Omega_{r}=\Omega_{br}$ and
$\Omega_{r}=0$, respectively. The dot dot dashed curves 6a and 6b represent
$t_{T}$ calculated for the star rotating with angular frequency
given by equation (\ref{e17a}) for values of binding energies larger
than that corresponding to the intersection of $t_{ev}$ and $t_{GW}$. For
smaller values of energies we have $t_{T}=t_{ev}$.
Note the slower evolution due to tides
for the faster rotation
rate. This results in gravitational radiation ultimately  controlling the
orbital evolution. 
All curves are calculated for $\mu=0.6,$ $\lambda =1.4,$ $\eta=5,$ $r_{1}=1$ and $M_{4}=1$.}
\label{fig1}
\end{figure}
\bigskip
\begin{figure}
\begin{center}
\includegraphics[width=9cm,angle=0]{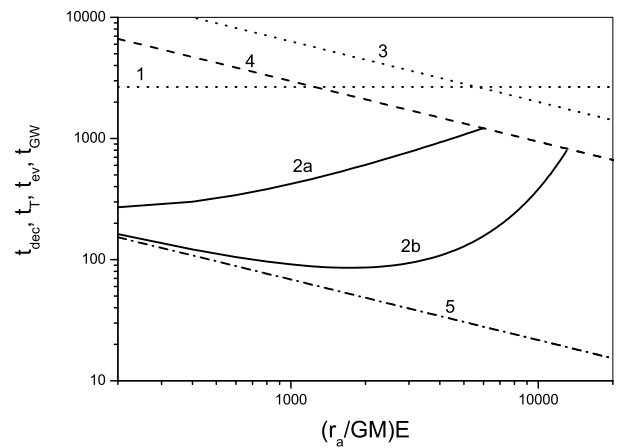}
\end{center}
\vspace{-1cm}
\caption{Same as Figure 1 but for $\eta=4$. Note that in this case we
  have $t_{ev}=t_{T}$ for the whole  range of binding energies shown.}
\label{fig2}
\end{figure}

Before going on to estimate of the rate of tidal capture and
circularisation we  use the results of the previous
sections to examine more closely the conditions
required for mode amplitude limitation due to gravitational wave emission
to enable safe circularisation with the possibility of tidal
heating, inflation and disruption avoided.

During the early evolutionary phases, 
subsequent to tidal capture, the star's orbit is highly eccentric, and the orbital 
angular momentum is approximately conserved during the orbital
evolution while the semi-major axis/binding energy as well 
as the angular velocity  of the star change with time (see eg. IP).
When tides dominate over the emission of gravitational waves induced by orbital motion, the
evolution time scale of the semi-major axis is given by 
equation (\ref{e18}) and we have $t_{ev}\approx t_{T}.$ When gravitational radiation
controls the orbital evolution, which happens at large enough orbital binding energy,
 we have $t_{ev}\approx t_{GW}$ where $t_{GW}$ is given by equation
(\ref{e22}). For a given value of the orbital angular momentum, the orbital
evolution time $t_{ev}$ depends on the orbital energy
$E$, and thus, on the star's semi-major axis $a$. This dependence
can be found from equations (\ref{e17a}),  (\ref{e18}),  (\ref{e19}),  (\ref{e18a}) and 
(\ref{e22}). For illustrative purposes we plot 
this dependence as well as that of other characteristic time-scales in
Figures 1 and 2 for a 'low density' white dwarf orbiting around a $10^{4}M_{\odot}$ black hole.
The cusp size is taken to be $1pc$. All dependencies are calculated in
assumption that the stochastic instability operates for the shown
range of binding energies.

In Figure 1 the characteristic time-scales corresponding
to a star with $\eta=5$ are shown. The horizontal line indicates the     
decay time scale of the stellar pulsations due to the emission of gravitational waves, $t_{dec}$,
given by equation (\ref{e14}). The inclined dotted line indicates the dependence
of the characteristic time $t_{GW}$ on ${\cal E}$ and the inclined dashed and dot-dashed lines
show the dependence of $t_{T}$ on ${\cal E}$ for $\Omega_{r}=\Omega_{br}=0.5\Omega_{*}$ 
 (dashed line) and $\Omega_{r}=0$ 
(dot-dashed line).  We recall that the timescale $t_T$ applies to the
evolutionary phase just after tidal capture which can exhibit stochastic instability. 
It may not be applicable to the later stages, when ${\cal E}$ is large, and
the conditions for additive impulsive energy inputs at pericentre passage
are not satisfied, see discussion below.

For a given value of ${\cal E},$ 
the dot-dashed line corresponding to a non rotating star, gives the smallest possible
value of $t_{T}=t_{min}$. For fixed  ${\cal E},$  stars with
values $t_{T}$ exceeding that given by the dashed line, would have angular 
velocities exceeding $\Omega_{br},$ and so would be disrupted. 
Therefore the dashed curve gives the largest allowed  value of  $t_{T}=t_{max}$.

The solid curves indicate the evolution of  $t_{ev}$
with $\Omega_{r}$ which is  given by equation (\ref{e17a}). The two different curves
correspond to the different assumptions about the value of the mode decay time due to
non-linear effects, $t_{nl}$. The curve which is uppermost 
at small values of ${\cal E}$ corresponds to the case a) where we assume that
$t_{nl}\approx t_{ev}$ while the other solid  curve corresponds to the  case b)
 for which  $t_{nl}$ is much less than both $t_{dec}$ and
$t_{ev}.$  One can see from  Figures \ref{fig1}  and \ref{fig2} that the behaviour of the
 characteristic time-scales
is rather similar for these two  cases. When $\cal E$ is very small,  rotation of the star
is small, and both curves are close to the dot-dashed line. Also, for small
values of $\cal E$ we have $t_{T}=t_{ev}$. 

When $\cal E$ increases, the star
is spun up by tides and $t_{T}$ gets larger.  When $\cal E$ attains a  value 
${\cal E}_{GW}\sim 10^{4}$ the solid  curves, 
which for  smaller values,  correspond to tidally driven evolution  cross the 
dotted line representing  evolution controlled by the gravitational radiation  time scale $t_{GW}$. For
larger values of ${\cal E},$ the orbital evolution is mainly determined by emission 
of gravitational waves, and we have  $t_{ev}= t_{GW} < t_{T}$. In this regime
the star continues to spin up by tides, and the tidal time scales shown
by dot dot dashed curves get larger with increasing 
of $\cal E.$  When the dot dot dashed curves cross the inclined dashed
line, the binding energies are equal to $E_{dis}$ given by equations
(\ref{en2}) and (\ref{en3}), and, according to
our criterion, the star is disrupted at energies $E\sim E_{dis}$. 

In Figure 2 we show the same quantities calculated for a star with a smaller value of
orbital angular momentum corresponding to $\eta=4$.
In this case the evolution is always controlled by tidal effects
and $t_{ev}=t_{T}$ for the whole  range of energies
shown ${\cal E} \le {\cal E}_{dis}$ and, accordingly, the the dot dot dashed curves coincide with the
solid curves.

As follows from this discussion and 
Figures {\ref {fig1}} and {\ref {fig2}}, the white dwarf has the possibility of surviving the tidal
evolution when the energy scale corresponding to the  onset of the stochastic instability
$E_{st}=GM/ (2a_{st})$ (see equation (\ref{e12})) is smaller than
$E_{dis}$ - the energy corresponding to  break-up rotation. 
When $E_{st} < E_{dis},$ disruption of the star is avoided during
the phase of stochastic instability. In this case, when $E > E_{st}$ the orbital evolution proceeds mainly
through emission of gravitational waves. Therefore, for our purposes it is very important
to establish under what conditions the inequality $E_{st} < E_{dis}$ holds.

As can be seen from Figures {\ref{fig1}} and {\ref{fig2}}, there are four different
possibilities: case a1) where $t_{nl}\approx t_{ev}$ and tides determine the orbital
evolution when the orbital energy of the star is close to $E_{dis}$;  case a2) where
 $t_{nl}\approx t_{ev}$ and the gravitational radiation determines the orbital
evolution;  case b1) when  $t_{nl}\ll t_{ev}$ and tides determine the evolution, and the case b2) where
 $t_{nl}\ll t_{ev}$ and  gravitational radiation controls the evolution.
Note that it is straightforward to see that
case b1) is always associated with disruption of the star. Indeed, in
this case the factor $t_{T}/t_{ev}$ entering equation (\ref{en3}) is
of order of unity. That means that the energy $E_{dis}$ is always small when
compared to $E_{st}$, for typical parameters of the system. Therefore,
the condition  $E_{st} < E_{dis}$ cannot be satisfied in this case and
the case b1) is not considered further. On the other hand, in the
opposite case b2) the factor  $t_{T}/t_{ev}$ can be large when
estimated at energies of the order of $E_{dis}$, see Figure 1, and the
condition  $E_{st} < E_{dis}$ can be fulfilled, see Appendix for details.  
In Appendix it is also shown that circularisation is possible for the cases a1) and a2)
provided that the orbital angular momentum of the star and, accordingly, $\eta$, is 
sufficiently large. Thus, circularisation can, in principal, be achieved for the cases
a1), a2) and b2).       
  
The condition $E_{st} < E_{dis}$ can also be reformulated in terms of 
semi-major axes. 
Using the expression (\ref{e17}) we can easily find the characteristic
semi-major axis $a_{dis}$ corresponding to the binding energy per unit mass
$E_{dis}$
\be a_{dis}= {r_{a}\over 2{\cal E}_{dis}}=5\cdot
10^{13}\left({t_{ev}\over t_{T}}\right)\left({t_{dec}\over t_{nl}+t_{dec}}\right)
\lambda\mu^{-1}M_{4}cm. \label{en5}\ee
As we discussed above when 
$a < a_{dis}$ and the condition for the stochastic instability
(\ref{e12}) holds, emission of gravitational waves/non linear effects 
cannot significantly reduce the energy and angular momentum  of the mode and the mode
 amplitude build-up persists. In such a situation a white dwarf is likely 
to undergo  intensive tidal heating which may possibly lead to 
disruption of the star. Therefore, the condition derived from
 (\ref{en5}) together
with corresponding conditions for onset of the stochastic instability
will be used in order to discriminate between orbital evolutionary 
tracks leading to disruption of the star and the tracks eventually 
leading to formation of close quasi-circular orbits around the
black hole.

When the non linear dissipation  time scale $t_{nl}$ is small (the cases b1 and b2), 
$a_{dis}$ does not depend on  $t_{nl}$ and we have 
\be a_{dis}= {r_{a}\over 2{\cal E}_{dis}}=5\cdot
10^{13}\left({t_{ev}\over t_{T}}\right)
\lambda\mu^{-1}M_{4}cm. \label{en6}\ee
When  $t_{nl}\approx t_{ev}$ (the cases a1 and a2) 
we can obtain an explicit expression for
$a_{dis}$ with the help of (\ref{e19}):
\be a_{dis}= {r_{a}\over 2{\cal E}_{dis}}=4.4\cdot
10^{16}\lambda^{8/3}\mu^{-2}M^{1/3}_{4}(\phi
\Psi)^{-2/3}cm. \label{e20}.\ee

The condition $a_{st} > a_{dis}$ (or $E_{st} < E_{dis}$) can be written as
a condition for the parameter $\eta $ to be larger than a certain value.
For our discussion below it is more convenient to use
the quantity $\phi(\eta )$ defined in equation (\ref {e7}) instead of $\eta$. 
Depending on which case is considered (i.e.  a1, a2 or b2),
the inequality  $a_{st} > a_{dis}$ can be written as a condition that the 
quantity  $\phi(\eta )$ is larger than its values
$\phi_{1}$, $\phi_{2}$ and $\phi_{3}$ corresponding to the equality $a_{st}= a_{dis}$
for the respective cases. 
These values are explicitly obtained in the Appendix, see equation
(\ref{e26}) for the case a1, equation (\ref{e29}) for the case a2 and equation 
(\ref{en8}) for the case b2.

\section{The rate of tidal capture by the black hole}\label{sec5}

In order to estimate the rate of tidal capture by the black hole  we assume
that the stars in the cusp are slowly rotating. Therefore,   
at the beginning of orbital circularisation the effects of rotation
are not important so we neglect it  when estimating  
the rate of  capture into circularising orbits,
 setting $\Psi=1$ in the expression
(\ref{e19}) for the time scale of orbital evolution.
Then  from the discussion  
in section 5 and data plotted in Figures \ref{fig1} and \ref{fig2}
 it follows that during this phase
orbital energy loss resulting from tides is more
important than that due to gravitational radiation.

The rate of  capture into circularising orbits, as a function of initial 
binding energy per unit mass  of the star, depends  on
whether the tidal time scale (\ref{e19}) is larger than
the orbital period $P_{orb}({\cal E})$. Equating 
(\ref{e19}) to (\ref{e3}) we find the binding energy per unit mass
 ${\cal E}_{1}$
where these two times are equal  to be given by
\be {\cal E}_{1}=0.7\phi^{-1}(\eta)({r_{a}\over R_{wd}}{m\over
M}), \label{e31} \ee
and we recall the  relation between $\eta$ and the 
$\Theta = J/J_{cone}$  that applies when  
the condition $M < M_{crit}$
 as given by  (\ref{e10}) is fulfilled is
$\Theta \approx \eta^{1/3}$.

\subsection{Balance of time scales for tidal evolution and
orbital evolution due to distant encounters}

At first we assume that ${\cal E}
> {\cal E}_{1}$ for a range of angular momenta of interest.
A star gets tidally captured by the black hole
when the characteristic time scale of evolution of its angular
momentum due to distant gravitational encounters with other stars
in the cluster
\be t_{J}\sim {J^{2}\over (\Delta j_{2})^{2}}P_{orb}, \label{e32}\ee
is larger than the tidal time scale $t_{T}$ (recall that 
the dispersion of angular momentum $j^{2}_{2}$ is given by equation
(\ref{e5})). Equating $t_{J}$ and $t_{T}$ we find the size of the "tidal
circularisation" loss cone in  angular momentum space as a
function of the dimensionless binding energy $\cal E$,
\be J_{T.C}=\sqrt {(\Delta j_{2})^{2}{t_{T}\over P_{orb}}}. \label{e33}\ee
Note that equation (\ref{e33}) defines $J_{T.C.}$ implicitly since this
quantity enters the right hand side through the dependence of
$t_{T}$ on $\eta.$ To deal with this we define
 $ \Theta_{T.C.} = J_{T.C}/ J_{T},$  where we recall that
$J_{T}=\sqrt{2GMr_{T}}.$ Then we have 
$ \Theta_{T.C.}= \eta^{1/3}$,
see equations (\ref{e8}) and (\ref{e9}).  Using the above,
an equation for $\Theta_{T.C.}$ may be obtained from
equations (\ref{e5}), (\ref{e18}) and  (\ref{e33}) in the form

\be \Theta_{T.C.}=1.14\mu^{-1/3}M_{4}^{5/6}\Lambda_{*}^{1/2}\phi^{1/2}(\Theta^{3}_{T.C.})
{\cal E}^{-3/4}, \label{e33a}\ee
where $\Lambda_{*}=\Lambda_{2}/8.5$.
It follows from equation (\ref{e18}) that in order to find $\Theta_{T.C.}$
and hence  
the size of the tidal  loss cone we   need to evaluate
the ratio $\Delta E_{T}/E$   when $J=J_{T.C.}.$ 
This  quantity grows from small values  as the orbital period increases
from small values, 
and accordingly  the  binding energy decreases, until it reaches unity 
 when $t_{J}=P_{orb}.$  At that point
equation (\ref{e31}) applies   as well as the condition  that
\be J^{2}_{T.C.}=(\Delta j_{2})^{2}. \label{e34}\ee
These  two equations define a critical energy $E_{crit}$.

 When $E < E_{crit}$ the tidal
evolution time scale is approximately equal to the orbital period 
and therefore the size
of the tidal circularisation loss cone is determined by equation
(\ref{e31}) which is, in this case, considered as an implicit equation
for $\Theta \equiv J_{T.C.}/J_{cone}$ for a given value of
$\tilde E$. Stars with these energies which have $J < J_{T.C.}$
calculated in this way  have relatively large energy changes
induced within a single orbit.

When $E > E_{crit}$ the number of stars tidally captured by the
black hole per unit of time, with typical energies of order of
some $E$, $\dot N_{T.C.}(E)$, can be estimated  to be
\be \dot N_{T.C.}(E) \sim N^0_{wd}(E,J_{T.C})(EJ_{T.C}/(\Lambda_1 t_{T}), \label{e35a}\ee
where we assume that the distribution function of the white dwarfs
is given by equation (\ref{e4})\footnote{Note an alternative form of
equation (\ref{e35a}) has already been provided
 in other studies, see e.g. Hopman $\&$
Alexander 2005, their equation (17) and references therein.   To obtain this we 
write $\dot N_{T.C.}(E)\sim
N^{0}_{wd}(E,J_{T.C})EJ_{T.C}/( t_{T}\Lambda_{1})=
N_{tot}  /(t_{r}  \Lambda_{1}) $, where
$N_{tot}\equiv N^{0}_{wd}(E, J_{circ})J_{circ}E$ characterises the total number
of stars having binding energies per unit mass 
 of order of $E$, we introduce the global
relaxation time scale $t_{r}\equiv J^{2}_{circ}P_{orb}/(\Delta j_{2})^{2}$
and use equations (\ref{e2}), (\ref{e4}) and (\ref{e33}).}.  
Using equations (\ref{e2}),
(\ref{e5}) and (\ref{e33}) we see that, in this case,
\be \dot N_{T.C.} \propto (\Delta j_{2})^{2}E \propto E^{-3/2} \label{e35}\ee
is mainly determined by stars with small energies. Note that we take into 
account the fact that $\Lambda_1$  depends
only logarithmically on the binding energy when deriving the 
relation (\ref{e35}), see equation (\ref{e33}).

For the case $E < E_{crit}$ it can be easily seen that $t_{J} <
t_{T}\approx P_{orb}$
for any value of $J$.
This situation is analogous to the full loss cone regime
in the theory of tidally disrupted stars (e.g. Frank $\&$ Rees
1976, Lightman $\&$ Shapiro 1977). In this case the rate of tidally captured stars can be
estimated as
\be \dot N_{T.C.} (E) \sim N^0_{wd}(E,J_{T.C})EJ_{T.C}/(  \Lambda_{1}P_{orb})\propto E, \label{e36}\ee
where here we take into account the fact that the size
of loss cone, now implicitly determined by equation (\ref{e31}), 
 depends logarithmically and hence  slowly  on $E$.

\subsection{Determination of the critical binding energy per unit
  mass}

 Thus, the rate of tidal circularisation is mainly determined by
 energies $E \sim E_{crit}$ (Novikov, Pethick $\&$ Polnarev 1992),
 and therefore it important to obtain
 an explicit expression for the value of $E_{crit}$. As we
 discussed above this value is determined by the condition that
the timescale for orbital evolution due to tides be equal to the orbital period
while being  on the boundary of the tidal loss cone.  
It is determined by equations (\ref{e31}) and
 (\ref{e34}) which can be rewritten in the  form
\be {\cal E}_{crit}\equiv {r_{a}\over GM}E_{crit}
=3\cdot 10^{5}\phi_{crit}^{-1}x, \label{e37} \ee
where $x=\mu r_{1}/( \lambda M_{4})$ and  $\phi_{crit} \equiv \phi(\Theta_{T.C.}^{3})$
is found from  equation
(\ref{e7}) after setting  $\eta=\Theta_{T.C.}^{3}$, together with
\be \Theta_{T.C.}^{2}={29\pi \over 40}\Lambda_{2}
q\left({r_{a}\over r_{T}}\right){\cal E}_{crit}^{-5/2}. \label{e38}\ee
Substituting ${\cal E}_{crit}$ from (\ref{e37}) to (\ref{e38}) we obtain
a single equation for $\eta$ and thence ${\cal E}_{crit}$ in the form
\be \phi_{crit} =1.7\cdot 10^{3}y^{3/5}\Theta_{T.C}^{4/5}, \label{e39}\ee
where 
\be y=\Lambda_{*}^{-2/3}\mu^{13/9}M_{4}^{-7/9}\lambda^{-1}r_{1}, \label{e40}\ee 
and $\Lambda_{*}=\Lambda_{2}/8.5$.
Using the explicit expression for $\phi(\eta)$ and the relation
$\eta=\Theta^{3}_{T.C.}$ we rewrite (\ref{e39}) in another form
\be \eta\approx 3.72+0.462\log \eta +0.219\log y. \label{e41}\ee
An approximate solution to equation (\ref{e41}) valid for the range
of interest of the parameter $y \sim 0.05-10$ with accuracy of order of $10$
per cent is
\be \eta \approx 4.3+0.22\log y, \label{e42}\ee
and
\be \phi_{crit}(\eta )  \approx 2\cdot 10^{3}y^{3/5}. \label{e43}\ee
As seen from equation (\ref{e43}) the size of the tidal
circularisation loss cone given by  $\Theta_{T.C.}=\eta^{1/3}$ corresponding to
the critical energy $E_{crit}$ is larger than
the size of the  tidal disruption
 loss cone given by  $\Theta=1$.

This is different from the situation where only the effect of emission of gravitational
waves is taken into account as a  mechanism for  changing  the orbital
energy. In that case the energy scale analogous $E_{crit}$ may also be
obtained from the   requirement that the orbital evolution time
$t_{ev}=t_{GW}$ be equal
to the orbital period on the 
boundary of the corresponding circularisation loss cone. 
Obviously, our estimate of the circularisation rate is valid only when 
the size of the circularisation loss cone is larger than the size of 'true'
loss cone, $J_{cone}$ corresponding to direct capture or disruption of stars by
the black hole.
However, for typical parameters expected for 
gravitational waves to be the dominant process of  orbital
evolution ($M \sim 10^{6}M_{\odot}$ and $r_{1}\sim 1$) the energy scale $E_{crit}$ 
is smaller than another energy scale $E_{cone}$ defined by the 
condition that the sizes of the circularisation and true loss cones coincide
at $E = E_{cone}$. When $E > E_{cone}$ the stars are mainly destroyed by the black hole
and the process of circularisation is strongly suppressed. 
In such a situation the capture rate is mainly
determined by energies $E\sim E_{cone}$ (Hopman $\&$
Alexander 2005).  Because of this difference, 
the tidal circularisation rate in our case,  (see equation (\ref{e47}) below)  differs
from the  estimate of the capture rate due to emission of
gravitational waves provided by Hopman $\&$
Alexander 2005, see also Novikov, Pethick $\& $ Polnarev 1992.

Equation (\ref{e37}) gives
\be {\cal E}_{crit} \approx 150 \left({x\over y^{3/5}}\right). \label{e44}\ee
We have explicitly
\be {\cal E}_{crit} \approx 150 \Lambda_{*}^{2/5}\left({r_{1}\over \lambda}\right)^{2/5}
\mu ^{2/15}M_{4}^{-8/15}. \label{e45}\ee
It is important to note that the critical energy ${\cal E}_{crit} \gg 1$
for our problem and therefore, the circularisation rate is mainly determined
by scales well within the cusp around the black hole. Also, equation
(\ref{e42}) tells that $\eta \sim 4$ as we have mentioned above.

\subsection{Rate of capture into circularising orbits}

The maximal rate of capture , $\dot N^{max}_{T.C}$ can be estimated
a little more accurately  as
\be \dot N^{max}_{T.C}\sim E_{crit}\int^{J_{T.C.}}_{J_{cone}}dJ
{N_{w.d}\over P_{orb}}, \label{e46}\ee
where $N_{w.d}$ is given by equation (\ref{e4}) with $E=E_{crit}$ and
$J_{cone}$ is defined in section 2.4. 
Let us recall that
in our problem  $J_{cone}=J_{T}$.
Integrating (\ref{e46}) we get
\be \dot N^{max}_{T.C.}={F\over \sqrt 2  \Lambda_{1}}\delta q^{-1}({r_{T}\over r_{a}})
{\cal E}_{crit} \Omega_{a}, \label{e47}\ee
where
\be F=(\Theta_{T.C.}^{2}(\ln \Theta_{T.C.}^{2}-1)+1)/4\approx 0.2, \label{e48}\ee
and we assume that $\Theta_{T.C}=\eta^{1/3}\approx 4^{1/3}$ in the last
equality.
In astrophysical units the expression (\ref{e47}) has the form
\be \dot N^{max}_{T.C.} \sim 2.5\cdot 10^{-8} \delta_{0.1}\Lambda_{*}^{2/3}
\Lambda_{8}^{-1}\mu^{-1/5}M_{4}^{13/10}\lambda^{3/5}
r_{0.1}^{-21/10}yr^{-1}. \label{e49}\ee
where $\delta_{0.1}=\delta/0.1$ and $\Lambda_{8}=\Lambda_{1}/8$.
Note that we use $r_{0.1}=r_a/(0.1pc)=10r_{1}$ in the last equation
assuming that a typical size of the radius of influence in 
globular clusters is of the order of $0.1pc$.

\subsection{Rate of capture of  stars that will circularise without disruption}
As  discussed in  section \ref{sec4}, 
a white dwarf has a possibility
of surviving  the process of circularisation, and accordingly,
 settling on a quasi-circular orbit only when a typical value of
the parameter $\eta$ (or equivalently the pericentre distance)
 is sufficiently large.
 The    tidally captured white dwarfs with formation rate 
  $\dot N^{max}_{T.C.}$ have a typical value of
$\eta$ given by equation (\ref{e42}) and, accordingly,
$\phi_{crit}(\eta )$ given by equation (\ref{e43}).

According to the  discussion  in  section \ref{sec4}  and the  Appendix,
such white dwarfs have a considerable survival probability,
for cases a1) and a2) corresponding to large $t_{nl,}$  when
\be \phi_{crit} > \phi_{max}\equiv \max (\phi_{1}, \phi_{2}), \label{e50}\ee
where $\phi_{1}$ and $\phi_{2}$ are determined by equations  
(\ref{e26}) and (\ref{e29}), respectively.
 For  case b2)  corresponding to small 
$t_{nl},$ for survival we require that
\be \phi_{crit} > \phi_{max}=\phi_{3} \label{e50n}\ee 
instead of  (\ref{e50}), where  $\phi_{3}$ is given by equation
(\ref{en8}). 

When the above conditions  {\it are not}  satisfied, 
the typical binding energy of  white dwarfs having 
a considerable probability of survival, $E_{s}$,   has to be larger than
the critical value 
 obtained from  equation (\ref{e45}) and so their capture rate
is reduced by a factor $(E_{s}/E_{crit})^{3/2}$
 when compared to that given 
by  equation
(\ref{e49}) (see equation (\ref{e35})).  Accordingly their
 capture rate 
 is given by
\be \dot N^{s}_{T.C.} \approx \dot
N^{max}_{T.C.}(E_{s}/E_{crit})^{-3/2}. \label{e51} \ee
The binding energy per unit mass,  $E_{s}$, is related to the quantity 
$\phi_{max}$ through the 'loss cone condition' (\ref{e33a}). Taking
into account the fact  that the  dependence of $\Theta_{T.C.}$ on
$E_{s}$ and $\phi_{max}$ is logarithmic, we
approximately have $E_{s}^{3/2} \propto \phi_{max}$, and therefore
equation (\ref{e51}) can be written as
\be \dot N^{s}_{T.C.} \approx \dot
N^{max}_{T.C.}(\phi_{crit}/\phi_{max}), \label{e52} \ee
and accordingly, the rate of formation is suppressed by the factor
$\phi_{crit}/\phi_{max} $ which is less than unity when conditions
(\ref{e50}) and (\ref{e50n}) are {\it not} satisfied.
As indicated below that is likely to be the situation for cases of interest.

We go on to obtain explicit expressions for the rate of capture of 
white dwarfs that may survive the circularisation process in that case. 
This rate 
will be referred to as  'the circularisation rate'.

First let us consider the case a) for which $t_{nl}\approx 
t_{ev}$. 
As follows from equation (\ref{e30}) $\phi_{2} > \phi_{1}$ when $M >
2\cdot 10^{3}M_{\odot}$. Since the systems containing sufficiently
massive black holes are likely to provide the bulk of the sources of
gravitational radiation, we assume hereafter that  $M >
2\cdot 10^{3}M_{\odot}$ and set, accordingly, $\phi_{max}=\phi_{2}$.
In this case the condition (\ref{e50}) leads to
\be r_{0.1} > 50\lambda^{2.1}\mu^{-3}M_{4}^{1.2}, \label{e53}\ee
where we use equations (\ref{e29}) and (\ref{e40}). In the 
systems of interest this condition is not satisfied, and  
indeed the circularisation rate is suppressed.

 We  use equations 
(\ref{e40}), (\ref{e43}), (\ref{e49}) and (\ref{e52}) to obtain the
the  suppressed capture  or circularisation rate in the form 
\be  \dot N^{s}_{T.C.} \approx 2.3\cdot 10^{-9}
\delta_{0.1}\mu^{1.6}M_{4}^{0.57}
\lambda^{-0.65} r_{0.1}^{-1.5}yr^{-1} ,\label{e54}\ee
where  hereafter  we set the logarithmic factors $\Lambda_{*}=\Lambda_{8}=1$.
Substituting values of $\lambda=1.4$ and $\mu =0.6$ appropriate
for a 'low density' white dwarf in (\ref{e54}), we obtain 
\be  \dot N^{s}_{T.C.} \approx 8.2\cdot 10^{-10}
\delta_{0.1}M_{4}^{0.57} r_{0.1}^{-1.5}yr^{-1},\label{e55}\ee
and in the case of the 'high density' white dwarf with
$\lambda =0.84$ and $\mu=1$, we  obtain 
\be  \dot N^{s}_{T.C.} \approx 2.6\cdot 10^{-9}
\delta_{0.1}M_{4}^{0.57} r_{0.1}^{-1.5}yr^{-1}.\label{e56}\ee

Now let us consider case b2) for which $t_{nl}\ll t_{ev}, t_{dec}$.
In this case the condition  (\ref{e50n}) can be expressed as 
\be r_{0.1} > 10\lambda^{3}\mu^{-135/45}M_{4}^{1/9}, \label{e56nn}\ee
where we have made use of equations (\ref{en8}) and (\ref{e43}). 
According to equation (\ref{e56nn}) the circularisation rate
is equal to the rate of tidal capture (\ref{e49}) when the cusp size
is sufficiently large.
For large cusp sizes the circularisation rate becomes
small due to a strong decrease of the capture rate as  it increases ( see equation (\ref{e49})).
Thus such systems are not likely to contribute significantly
to the total number of gravitational wave sources.
 Therefore, we assume hereafter that the inequality
 (\ref{e56nn}) does not apply and multiply
(\ref{e49}) by the suppression factor $(\phi_{crit}/\phi_{3})$ in
order to obtain the circularisation rate. 
Thus we have
\be \dot   N^{s}_{T.C.} \approx 6.2\cdot 10^{-9}
\delta_{0.1}\mu^{122/75}M_{4}^{37/30}\lambda^{-6/5}r_{0.1}^{-1.5}yr^{-1}.\label{e54n}\ee
Substituting values appropriate to the two types of white dwarf  in
(\ref{e54n}), we  obtain 
for the 'low density' case
\be \dot   N^{s}_{T.C.} \approx 1.8\cdot 10^{-9}
\delta_{0.1}M_{4}^{1.2}r_{0.1}^{-1.5}yr^{-1},     \label{e55n}\ee
and
\be  N^{s}_{T.C.} \approx 7.7\cdot 10^{-9}
\delta_{0.1}M_{4}^{1.2}r_{0.1}^{-1.5}yr^{-1}     \label{e56n}\ee
for the 'high density case'.

A comparison of equation (\ref{e49}) with equations (\ref{e54}),
(\ref{e55}), (\ref{e56}),  (\ref{e54n}),
(\ref{e55n}), (\ref{e56n}) shows that our criterion of survival
of the white dwarfs during the process of circularisation typically results 
in an order of magnitude decrease of the circularisation rate.
However note that for case b2) when $t_{nl}$ is small,
the suppression is not significant for the 'high density' white dwarfs..

\section{Probability of detection of a source of gravitational
radiation}\label{sec6}

The probability of formation of  observable sources of gravitational waves
is determined by several important factors such as the rate of formation
of white dwarfs on circularising orbits 
calculated above, the subsequent orbital evolution of the white dwarfs,
 the abundance
of globular clusters having sufficiently massive black holes
and the properties of a gravitational wave antenna 
receiving the signal.  In what follows we  make 
a rough estimate of the probability assuming that the gravitational
wave antenna has characteristics close to what has been indicated  for
the future LISA space borne gravitational wave antenna. 

LISA has its maximal sensitivity in the frequency range $\sim
10^{-2}Hz$   for which the dimensionless amplitude of the gravitational 
waves, $h$, can be as small as
\be h \approx 10^{-24}\sqrt{{1yr\over T_{obs}}}, \label{e57}\ee
where $T_{obs}$ is the time of observation,
see e.g. Grishchuk, Lipunov, Postnov, Prokhorov and
Sathyaprakash 2001 and references therein. 
Therefore, for simplicity, we assume that the white dwarf has  an
orbital period $P_{orb}\approx 100s$ 
when emitting the gravitational waves to be received by the antenna.
The time $T_{obs}$ can be smaller than or of the same 
order as the orbital evolution
time $T_{ev}$ calculated for an orbital period $\sim 100s$, 
and we assume hereafter that these are equal to each other,
$T_{obs}=T_{ev}$. In order to estimate $T_{ev}$ we  need to specify
a typical value of eccentricity of the orbit, $e$, which, in turn,
is determined by whether the orbital evolution is governed by tides
or by the emission of the gravitational waves. 
For such  small orbital periods the evolution is likely to be
determined by the latter effect and we can
use the results obtained by Peters $\&$ Mathews 1963 and Peters
1964 when estimating $e$.  To do that we proceed as follows. First
we estimate the orbital semi-major axis for  an orbit with
$P_{orb}=100s$  from Kepler's law as 
\be  a \approx 4.7 \mu^{1/3}\lambda^{-1}r_{T}. \label{e58}\ee
Next we use the dependence of the orbital angular momentum on 
$e$ obtained by Peters, 1964
\be J=0.94H(e)e^{6/19}\eta_{t}^{1/3}J_{T}, \label{e59}\ee
where $H(e)=(1+{121\over 304}e^{2})^{435/2299}$. Equation (\ref{e59}) 
implies that when $e \sim 1$
the orbital angular momentum
is equal to $J=\eta_{t}^{1/3}J_{T}$. 
Since a typical value of $\eta $ allowing survival 
 during the process of tidal
circularisation, according to section 5 and Appendix,
is slightly larger than the   typical value
$\eta\approx 4$, we set $\eta_{t}=5$ hereafter. Now we can
equate the standard expression $J=\sqrt{GMa(1-e^{2})}$ to
what is given in equation (\ref{e59}), and using equation
(\ref{e58}) with $\eta_{t}=5$ 
obtain an implicit equation for the eccentricity $e$
\be 0.9\mu^{1/3}\lambda^{-1}(1-e^{2})=H^{2}(e)e^{12/19}. 
\label{e60}\ee
From this equation we  obtain $e\approx 0.56$ and $e\approx 0.31$ 
for the high/low density white dwarfs, respectively.
The evolution time can be expressed
in the form
\be T_{ev}=1.4F(e)M_{4}^{-2/3}P_{2}^{8/3}yr, \label{e61}\ee
where $P_{2}=P_{orb}/100s$ and the function $F(e)$ specifies 
the dependence of $T_{ev}$ on $e$, see Peters 1964. 
We have
$F(e=0)=1$, $F(e=0.31)\approx 0.8$ and $F(e=0.56)\approx 0.2$.
Note, that an analogous expression for $T_{ev}$ 
given in Ivanov 2002 in his
equation (79) contains an error in last expression on the right
hand side and must be multiplied on
additional factor $M/(10^{6}M_{\odot})$. As a result, although
his estimate of a probability of finding a source of gravitational 
waves  in galactic centres containing black holes of masses
$\sim 10^{6}M_{\odot}$ (equation (80) of his paper) 
is numerically correct for the parameters
he uses, it has a different scaling with the black hole mass.
It must be multiplied by  an additional factor $M/(10^{6}M_{\odot})$
which results in smaller values of the probability for 
black hole  masses in the range $\sim 10^{4}-10^{6}M_{\odot}$.

Now, substituting (\ref{e61}) in (\ref{e57}) we obtain 
minimal values of the amplitude $h$, $h_{min}$, which can
be detected by LISA
\be h^{(l)}_{min}\approx 10^{-24}M_{4}^{1/3}, \quad h^{(h)}_{min}\approx 5\cdot
10^{-24}M_{4}^{1/3}, \label{e62}\ee
where $h_{min}^{(l,h)}$ stands for the minimal amplitudes for
the cases of low/high density white dwarfs, respectively, and
we set $P_{2}=1$. 

In order to find the total volume of space, available 
for detection of the signals by LISA, we should relate the quantities
$h_{min}^{(l,h)}$ to the distance from a source, $R_{obs}$. For that
the amplitude $h$ of radiation emitted by the source should be known
as a function of distance and of parameters of the source. 
The appropriate relation has been obtained,
e.g., by Nelemans, Yungelson and Portegies Zwart 2001. In our units
it has the form
\be h(R_{obs}) \approx 5\cdot 10^{-21}\mu f(n,e)M_{4}^{2/3}P_{2}^{-2/3}{1Mpc\over
R_{obs}}, \label{e63}\ee
where the function $f(n,e)$ contains information about the amplitudes
 associated with harmonics  of different order $n$. It is related to
the function $g(n,e)$ calculated by Peters $\&$ Mathews 1963
  through $f(n,e)=\sqrt{g(n,e)}/n$
\footnote{Note a misprint in Appendix of Peters $\&$ Mathews 1963 in
their expression for $g(n,e)$. The fifth term in the braces must
contain the factor $(4/e)^2$ contrary to the factor   $(4/e^2)^2$ as
in the text.}. We have $f(n,e=0.56)\approx 0.2$ for 
several harmonics with $n=1,2,3$ and  
$f(n=2,e=0.31)\approx 0.4$. 
Now we can equate $h(R_{obs})$ to $h^{(1,2)}_{min}$ and find the
maximal distance to a source which can be, in principal, detected
by LISA
\be R_{obs}^{(l)}=1200M_{4}^{1/3}Mpc,  \quad
R_{obs}^{(h)}=200M_{4}^{1/3}Mpc, \label{e64} \ee
where the upper indices $(l,h)$ stand for the cases of low/high
density white dwarfs hereafter.   

In order to find the total number of  potential sources, $N_{tot}$, 
we use an estimate obtained by Portegies Zwart $\&$ McMillan 2000 for 
the average number of globular clusters in the Universe,
$n_{G.C.}\approx 3Mpc^{-3}$ for the value of the Hubble constant
$H=70km\cdot s^{-1}\cdot Mpc^{-1}$
\footnote{Note that the average number of globular clusters is meaningful only when
$R_{obs} > 70Mpc$, where $70Mpc$ is a typical distance between the elements
of the Large Scale Structure in the Universe.}. We obtain from equation (\ref{e64})
 $$ \hspace{-2.2cm} N_{tot}^{(l)}={4\pi\over 3}n_{G.C.}\Delta (R^{(1)}_{obs})^{3}\approx
2\cdot 10^{10}\Delta M_{4}, \quad $$ 
\be N_{tot}^{(h)}
 \approx  10^{8}\Delta
M_{4}, \label{e65}\ee
where $\Delta $ is the fraction of globular clusters having black
holes with masses $\sim M_{4}$. The total probabilities of finding of 
a source, $Pr^{(1,2)}$, can be estimated as products of the circularisation rates
given by equations (\ref{e55}) and (\ref{e56}) for  case a) and 
equations (\ref{e55n}) and (\ref{e56n}) for case b), the total numbers
of the sources available and the operational time of the LISA mission taken
to be equal to $2yr$.

For case a) with  $t_{nl}\approx t_{ev},$ using equations
(\ref{e55}), (\ref{e56}) and (\ref{e65}) we have
\be Pr^{(l)}\approx  30\alpha , 
\quad Pr^{(h)}\approx 0.4\alpha, \label{e66}\ee
where 
\be \alpha=\Delta \delta_{0.1}M_{4}^{1.57}r_{0.1}^{-1.5}. \label{e67}\ee
It is seen from equation (\ref{e66}) that the high density white dwarfs of
Sirius B type give rather small probability of detection. On the other 
hand, the 'ordinary' white dwarfs could provide, in principal, 
a source of gravitational waves provided that our criterion of white
dwarfs survival is valid and that
\be \alpha > 1/30. \label{e68}\ee
It is interesting reformulate the criterion (\ref{e68})
in term of a typical velocity of stars near the radius of influence
of the black hole, $\sigma=\sqrt{GM/r_{a}}$. We have from equations 
(\ref{e67}) and (\ref{e68})
\be \Delta > {1\over 30 \delta_{0.1}}\left({ 20km/s \over \sigma }\right)^{3}, \label{e69}\ee
where we neglect a very slow dependence on the black hole mass, 
$\alpha(\sigma, M) \propto M^{0.07}$. Of course, the black hole mass
 must be sufficiently large for the  validity of the  assumptions
leading to equation (\ref{e68}). 

For case b) with $t_{nl}\ll t_{ev}, t_{dec}$ we proceed in a similar manner
but use equations (\ref{e55n}) and (\ref{e56n}) instead of  equations
(\ref{e55}) and (\ref{e56}), respectively. We find
\be Pr^{(l)}\approx  70\alpha , 
\quad Pr^{(h)}\approx 1.5\alpha, \label{e66n}\ee
where
\be \alpha=\Delta \delta_{0.1}M_{4}^{2.2}r_{0.1}^{-1.5}. \label{e67n}\ee 
and
\be \Delta > {1\over 70 \delta_{0.1}}\left({ 20km/s \over \sigma }\right)^{3}M_{4}^{-0.7}, \label{e69n}\ee
instead of equation  (\ref{e69}) for a  'typical low density'  white dwarf.
 
As for case a) 'typical' white dwarfs  have a much larger probability of detection. 
However, for case b) this probability depends explicitly on the black hole mass 
when the cusp size is expresses in terms of the typical stellar
velocity dispersion  $\sigma$. As follows  from equations (\ref{e66n}), (\ref{e67n}) and (\ref{e69n}) when
$M_{4}\sim 1$ and $\sigma \sim 20km/s$ this probability is about
two times larger than that corresponding to case a). But the probability ratio  decreases
with the black hole mass  if  $\sigma $
is kept fixed.

\section{Discussion }\label{sec7}
\noindent
In this paper we have   given a qualitative analysis of the problem of tidal
circularisation of white dwarfs in globular clusters containing black
holes with masses in the range $10^{3}-10^{4}M_{\odot}$ and estimated 
the rate of production of
white dwarfs that  begin to  circularise  their orbits
 for   specified  parameters for the
cluster, white dwarf and black hole, see equation (\ref{e49}). We also
proposed a simple criterion for 'survival' of a white dwarf during 
the stage of orbital circularisation, and found  production rates for 
 white dwarfs, which can, in principal, settle down in a
quasi-circular orbit thus forming sources of gravitational radiation,
see equations (\ref{e54}-\ref{e56}) and (\ref{e54n}-\ref{e56n}). 
These rates are order of magnitude
smaller than that  given by equation (\ref{e49}). We made a simple
estimate of the probability of detection of  systems containing  
white dwarfs on quasi-circular orbits and found that these systems can,
in principle, be detected by LISA provided that the globular clusters
containing  black holes are sufficiently abundant and their stellar  velocity
dispersion near the radius of influence of the black hole is
sufficiently large, see equation (\ref{e69}) and (\ref{e69n}).

Our results should be treated  with  caution.  
Many processes occurring during the formation of  circularising stars
and during ongoing tidal circularisation  depend strongly  on the  parameters 
of the problem and this leads to uncertainties in the results. 
 Let us  consider some of them.

\noindent 1) In a quantitative approach a realistic model of a white dwarf 
 with non-zero temperature should be used. The eigen spectrum of
perturbations should include the gravity/inertial modes. The
realistic model should also take into account different channels
of mode decay and a self-consistent change of 
structure of the white dwarf
as a result of radiative cooling/heating due to dissipation of the mode
energy.      

\noindent 2) Our results indicate that the stars can achieve very high rotation
rates during the process of ongoing tidal circularisation. The
possibility of survival of white dwarfs during this stage critically 
depends on tidal exchange of energy between the orbit and 
the modes of a fast rotating star. It  will be very important  
to generalise well known results of the theory of dynamic 
tides in non-rotating/slowly rotating stars to the 
case of fast rotators.

\noindent 3) As follows from this discussion, our results depend
on the rate of mode energy and angular momentum transfer to the star.
Possible channels for this transfer, especially 
through dissipation resulting from non linear effects should be  
further investigated. 

\noindent 4) The dependence of the circularisation rate on the cusp size 
and black hole mass should be checked with
numerical N-body simulations. Also, a more realistic distribution
of the white dwarfs over the cluster as well as a change of 
their relative fraction due to stellar evolution should be
accounted for in a numerical study. 

\noindent 5) We have adopted a rather   simplistic criterion for detection 
of the gravitational waves produced by the white dwarf orbiting
close to a black hole, assuming that the amplitude of the 
signal is equal to the noise amplitude of the  detector. Higher
S/N detection threshold would result in a much smaller probability of detection.
On the other hand, a more sophisticated strategy of  signal  
processing could improve our estimates.

We have shown that during the first stage of orbital circularisation
the orbital energy  transferred to the fundamental mode of 
stellar pulsations may be radiated away by gravitational waves. That means
that the globular clusters may contain sources of gravitational
radiation with a typical frequency of the order of  the eigen frequency
of the fundamental quadrupole mode, of the order of a few Hz. 
 The possibility of detection of such sources needs further 
investigation. 
  
Finally we comment that in a recent paper
 Baumgardt, Hopman, Portegies Zwart, and
Makino 2006 have calculated the tidal capture rates for stars of 
different types with help of N-body simulations, during 
first 12Myrs of evolution of the system. They have considered 
black holes with masses in the range 
$\sim 10^{3}-4\cdot 10^{3}M_{\odot}$. Giant stars and low mass
main sequence stars with mass $< 0.4M_{\odot}$ have been modelled
as $n=1.5$ polytropes. The low mass stars seem to be absent in the
cusp due to the effect of mass segregation operating 
during formation of the cusp.  
However, their capture rate of red giants can be compared
with what is given by equation (\ref{e49}). They have obtained a capture
rate for  red giants of the order of a few events per  run. 
Assuming that a typical red giant has a radius $\sim
10^{3}R_{\odot}$, mass $\sim 10M_{\odot}$,  taking the black hole
mass and the cusp size  to be equal to $3\cdot 10^{3}M_{\odot}$ and to
$0.1pc$, respectively, and  substituting these values into
equation (\ref{e49}), we obtain 
$\dot N_{R.G.}\sim 3\cdot 10^{-5}\delta_{R.G.}yr^{-1}$, where 
$\delta_{R.G.}$ is the number fraction of the red giants in the 
cusp. Our results appears to be
in full agreement with the results of N-body simulations provided 
that $\delta_{R.G.}\sim 1/100$. 

\section*{Acknowledgements }
\noindent
We are grateful to A. G. Polnarev for fruitful discussions 
to A. G. Doroshkevich for useful remarks and to the referee for valuable comments. 
PBI has been supported
in part by RFBR grants 04-02-17444 and 07-02-00886.


\begin{appendix}

\section{Conditions for the transition from  stochastic instability}\label{3.2}

As indicated in section \ref{stocha}, there is a phase of evolution
at large semi-major axis subsequent to tidal capture
where the internal pulsation mode energy may increase in a stochastic manner.
This requires the semi-major axis to exceed $a_{st}$ above which
the pulsation mode does not maintain phase coherence
between successive pericentre passages.

Here we obtain an explicit expression for the
semi-major axis $a_{st}$ defined  through equation (\ref{e12}). 
At first let us
assume that $\Delta E_{T} > \Delta E_{GW}$ during the orbital
evolution. This situation is illustrated in Figure 2.
In this case we substitute equation (\ref{e6}) in
equation (\ref{e12}) and obtain
\be a_{st}\equiv a^{(1)}_{st}\approx 1.15\cdot 10^{11}\lambda \mu^{-3/5} M_{4}^{3/5}
(\phi \Psi)^{2/5}cm. \label{e23}\ee 

Now let us consider the    orbital evolution  when $\Delta E_{T} < \Delta
E_{GW}$. This situation is illustrated in Figure 1.
In this case the semi-major corresponding to the onset
of the stochastic instability, $a_{st} \equiv a_{st}^{(2)}$ 
can be obtained by substitution of equation (\ref{e21}) 
in equation (\ref{e12}) with the result
\be a^{(2)}_{st}\approx 1.7\cdot 10^{12}\eta^{14/15}\lambda^{2}
\mu^{-16/15}M_{4}^{1/15}cm. \label{e27}\ee

Just after tidal capture occurs the orbital evolution is primarily due to
tides. The orbit decays until  $a_{st}$ is reached but at this point
it is possible that the orbital evolution   is governed by gravitational
radiation. As indicated in section \ref{sec4}, when the orbital
semi-major axis is of the order of $a_{st}$
there are  three non trivial
possibilities:
 case a1) where $t_{nl}\approx t_{ev}$ and tides determine the orbital
evolution;  case a2) where
 $t_{nl}\approx t_{ev}$ and  gravitational radiation determines the orbital
evolution; and   case b2) where
 $t_{nl}\ll t_{ev}$ and  gravitational radiation controls the orbital 
 evolution. Each of these is considered in more detail below.

\subsection{The case a1: tides dominate
and $t_{nl}$ is large}\label{3.2n}

As indicated in sections 4 and 5, when $t_{nl}\sim t_{ev}$
a white dwarf has a possibility of surviving
 the  process of tidal circularisation only if it is cooled 
efficiently by emission of gravitational waves   for
 values of  its semi-major axes $>a^{(1)}_{st}$. Therefore, in order
to reach a quasi-circular orbit, the white dwarf must have orbital parameters 
such that $a^{(1)}_{st} > a_{dis}$  where $a_{dis}$ is given by
equation (\ref{e20}). This inequality leads to 
an inequality   of the form $\eta > \eta_{1}$ ( or equivalently
the fixed pericentre distance must exceed a certain value),
 where $\eta_{1}$
is defined by the condition $a^{(1)}_{st}=a_{dis}$. In general, $\eta_{1}$
is a monotonic 
 function of the stellar rotation rate with smaller values of $\eta_{1}$
corresponding to larger values of $\Omega_{r}.$ This is because tides
weaken with increasing  $\Omega_{r}$ enabling safe circularisation
starting with smaller pericentre distances.

  Thus in order to take into account all possible evolutionary
tracks leading to formation of circular orbits, we calculate the lower
boundary of $\eta_{1}$ assuming that 
$\Omega_{r}=\Omega_{br}\approx 0.5 \Omega_{*}$. In this case the 
factor $(\phi \Psi)$ entering (\ref{e23}) takes
the form
\be (\phi \Psi)\approx {1\over \eta}\exp{(3.68\eta -2.74)}. \label{e24}\ee
Substituting equation (\ref{e24}) in equation (\ref{e23}) and equating
(\ref{e20}) and (\ref{e23}), we obtain
\be \eta_{1}\approx 4+0.27(\ln \eta_{1} +\ln Z_{1}), \label{e25}\ee
where $Z_{1}=\lambda^{25/16}\mu^{-21/16}M_{4}^{-1/4}$.

In   the subsequent discussion we  
 use the quantity $\phi_{1}\equiv \phi(\eta_1)$ defined in equation (\ref{e7})
rather than $\eta_{1}$. This quantity can be approximately represented
in the form
\be \phi_{1}\approx 2.4\cdot 10^{3}Z_{1}^{0.7}\approx 2.4\cdot
10^{3}\lambda^{1.1}\mu^{-0.92}M_{4}^{-0.175}, \label{e26}\ee
with the difference between this power-law fit and the solution of
equation (\ref{e25}) being less than or of the order of 5 per cent 
for all interesting values of $Z_{1}$. 

\subsection{The case a2: gravitational waves dominate
and $t_{nl}$ is large}\label{3.3}

Now let us assume that gravitational waves determine the orbital evolution 
when a rapidly rotating white dwarf has its semi-major axis near to
$a_{dis}$ and $t_{nl}\approx t_{ev}\approx t_{GW}$, see Figure 1.
Proceeding as in the preceding section,  we 
 obtain an  equation for the critical value of $\eta_{2}$
that is analogous to equation (\ref{e25}) in the form
\be \eta_{2}\approx 4.88 -0.11\ln \eta_{2} +0.27\ln Z_{2},
\label{e28}\ee
where $Z_{2}=\lambda \mu^{-7/5}M_{4}^{2/5}$. The
corresponding value of $\phi_{2}\equiv \phi(\eta_{2})$ can
also be, with a high accuracy, represented in the power-law 
form
\be \phi_{2}\approx 5.4\cdot 10^{3}Z_{2}^{13/20}\approx
5.4\cdot 10^{3}\lambda^{0.65}\mu^{-0.91}M_{4}^{0.26}. \label{e29}\ee
      
The value of the parameter $\eta $ and  thus 
$\phi(\eta )$ which divides, according to our criterion, 
 orbits  leading  to a safe circularisation
from  orbits leading to a possible disruption of the white dwarf,
 is obtained from  the  condition $\phi(\eta ) > max(\phi_{1}, 
\phi_{2}).$   We note that the condition for gravitational
radiation to dominate,  $\phi_{2} > \phi_{1},$
  leads to the approximate condition on the black hole mass that 
\be M_{4} > M_{*}=0.155\lambda, \label{e30}\ee
where we have neglected a very   weak dependence of $M_{*}$ on $\mu$
of the form $M_{*}\propto \mu^{-0.02},$ the latter quantity being
replaced by unity.

The inequality (\ref{e30}) implies that for a sufficiently large
values of the black hole mass and the rotation of the white
dwarf the orbital evolution is mainly determined by emission 
of gravitational waves when the semi-major of the orbit 
is close to the characteristic semi-major axis  corresponding to the onset
of  stochastic instability. For the parameters corresponding
to a 'typical' white dwarf with $\mu=0.6$ and $\lambda =1.4$, 
we have $M_{*}\approx 0.22$,  
and  for a dense white dwarf of Sirius B type with $\mu=1$
and $\lambda=0.84$,  we have $M_{*}=0.13$. Both characteristic values
of $M_{*}$ are smaller than  the black hole masses of 
systems expected to produce a large amount of gravitational 
radiation during the last stages of orbital evolution of the star.
Therefore, the condition $\eta > \eta_{2}$ is more important for
our purposes.       

\subsection{The case b2: gravitational waves dominate and  $t_{nl}$ is small}\label{3.4}

Now let us consider the case when  $t_{nl}\ll t_{ev}, t_{dec}$. The discussion
proceeds in a similar way to the previous case  but now
$a_{dis}$ is given by equation (\ref{en6}). As we have mentioned above
it can then be shown
that  gravitational waves dominate the orbital evolution 
when the semi-major axis is $\sim a_{dis}$ provided that  
the black hole mass exceeds $\sim 10^{3}M_{\odot}$, 
and accordingly, $t_{ev}\approx t_{GW}$.

Therefore in order to obtain an equation for the parameter
 $\eta \equiv \eta_{3}$ separating orbits leading to safe circularisation from those
 leading to  disruption of the star, we equate expressions (\ref{en6}) and  (\ref{e27}),
and use equations (\ref{e6}), (\ref{e21}), (\ref{e22}) and  (\ref{e24}), 
to obtain
\be \eta_{3}=3.5+0.65\ln \eta_3 +0.27\ln Z_{3},   \label{en7}\ee
where $Z_{3}=\lambda^{3/2}\mu^{-6/5}M_{4}^{-8/15}$.   
An approximate solution to this equation can be   found from
\be \phi_{3}\equiv \phi(\eta_{3})\approx 
2\cdot 10^{3}\lambda^{6/5}\mu^{-24/25}M_{4}^{-32/75}. \label{en8}\ee

\end{appendix}

{}

\clearpage

\centerline {\bf        }
\vspace{1 cm}

\clearpage

\end{document}